\documentclass{aa}  
\usepackage{graphicx}
\usepackage{txfonts}
\usepackage{comment}
\usepackage{xcolor}

\begin{document}

   \title{The Close AGN Reference Survey (CARS)}
  \subtitle{A parsec scale multi-phase outflow in the super-Eddington NLS1 Mrk~1044}


   \author{N.~Winkel\inst{\ref{MPIA}}
            \and
          B.~Husemann\inst{\ref{MPIA}}
            \and
          M.~Singha\inst{\ref{Winnipeg}}
            \and
          V.~N.~Bennert\inst{\ref{CalPoly}}
            \and
          F.~Combes\inst{\ref{Paris}}
            \and
          T.~A.~Davis\inst{\ref{Cardiff}}
            \and
          M.~Gaspari\inst{\ref{Princeton}}
            \and
          K.~Jahnke\inst{\ref{MPIA}}
            \and
          R.~McElroy\inst{\ref{Queensland}}
          \and 
          C.~P.~O'Dea\inst{\ref{Winnipeg}}
          \and 
          M.~A.~P\'erez-Torres\inst{\ref{Granada}, \ref{Zaragoza}, \ref{Cyprus}}
          }

    \institute{
      Max-Planck-Institut f\"ur Astronomie, K\"onigstuhl 17, D-69117 Heidelberg, Germany, \email{winkel@mpia.de} \label{MPIA}
      \and
      Department of Physics \& Astronomy, University of Manitoba, Winnipeg, MB R3T 2N2, Canada \label{Winnipeg}
      \and
      Physics Department, California Polytechnic State University, San Luis Obispo, CA 93407, USA \label{CalPoly}
      \and
      LERMA, Observatoire de Paris, PSL Univ., Coll\`ege de France, CNRS, Sorbonne Univ., Paris, France \label{Paris}
      \and
      Cardiff Hub for Astrophysics Research \&\ Technology, School of Physics \& Astronomy, Cardiff University, CF24 3AA, UK \label{Cardiff}
      \and 
      Department of Astrophysical Sciences, Princeton University, 4 Ivy Lane, Princeton, NJ 08544-1001, USA \label{Princeton}
      \and
      School of Mathematics and Physics, The University of Queensland, St. Lucia, QLD 4072, Australia \label{Queensland}
      \and
      Instituto de Astrof\'isica de Andaluc\'ia (IAA-CSIC),Glorieta de la Astronom\'ia s/n, E-18008 Granada, Spain \label{Granada}
      \and
      Facultad de Ciencias, Universidad de Zaragoza, Pedro Cerbuna 12, E-50009 Zaragoza, Spain \label{Zaragoza}
      \and 
      School of Sciences, European University Cyprus, Diogenes street, Engomi, 1516 Nicosia, Cyprus \label{Cyprus}
      }

 
  \abstract
   {The interaction between Active Galactic Nuclei (AGN) and their host galaxies is scarcely resolved. Narrow-line~Seyfert~1 (NLS1) galaxies are believed to represent AGN at early stages of their evolution and allow to observe AGN feeding and feedback processes at high accretion rates.
   }
   {
   We aim to constrain the properties of the ionised gas outflow in Mrk~1044, a nearby super-Eddington accreting NLS1. Based on the outflow energetics and the associated timescales, we estimate the outflow's future impact on the ongoing host galaxy star formation on different spatial scales.
   }
   {
   We apply a spectroastrometric analysis to VLT MUSE NFM-AO observations of Mrk~1044's nucleus. This allows us to map two ionised gas outflows traced by [\ion{O}{iii}] which have velocities of $-560 \pm 20\,{\rm km\:s}^{-1}$ and $-144 \pm 5 \,{\rm km\:s}^{-1}$.
   Furthermore, we use an archival HST/STIS spectrum to identify two Ly-$\alpha$ absorbing components that escape from the centre with approximately twice the velocity of the ionised gas components.
   }
   {Both [\ion{O}{iii}] outflows are spatially unresolved and located close to the AGN ($< 1\,{\rm pc}$). They have gas densities higher than $10^5 \,{\rm cm}^{-3}$, which implies that the BPT diagnostic cannot be used to constrain the underlying ionisation mechanism. 
   We explore whether an expanding shell model can describe the velocity structure of Mrk~1044's multi-phase outflow. 
   We find an additional ionised gas outflowing component that is spatially resolved. It has a velocity of $-211 \pm 22 \,{\rm km\:s}^{-1}$ and projected size of \mbox{$4.6 \pm 0.6 \,{\rm pc}$}.
   A kinematic analysis suggests that significant turbulence may be present in the ISM around the nucleus, which may lead to a condensation rain, potentially explaining the efficient feeding of Mrk~1044's AGN.
   Within the innermost 0.5$\arcsec$ (160$\,{\rm pc}$) we detect modest star formation hidden by the beam-smeared emission from the outflow. 
   }
   {We estimate that the multi-phase outflow has been launched $< 10^4 \,{\rm yrs}$ ago. Together with the star formation in the vicinity of the nucleus, this suggests that Mrk~1044's AGN phase set on recently.
    The outflow carries enough mass and energy to impact the host galaxy star formation on different spatial scales, highlighting the complexity of the AGN feeding and feedback cycle in its early stages.
   }
   
   \keywords{galaxies: kinematics and dynamics - galaxies: ISM - galaxies: active - galaxies: Seyfert - quasars: supermassive black holes - quasars:individual: Mrk 1044}

   \maketitle
%

\section{Introduction}

Supermassive black holes (SMBHs) in the hearts of galaxies are believed to interact with their host galaxy. The enormous amount of energy released by an active galactic nucleus (AGN) is injected into the interstellar medium (ISM) where it effectively delays the gas cooling. This deprives the galaxy of its cold gas reservoir \citep{Zubovas&King:2012} and eventually leads to star formation (SF) quenching, an effect that is referred to as `negative feedback'. The AGN-induced negative feedback is an important ingredient to reproduce galaxy properties in cosmological hydrodynamical simulations \citep[e.g.][]{Crain:2015, Dubois:2016, Weinberger:2017} and semi-analytic models \citep[e.g.][]{Somerville:2008, Fontanot:2020}.
In contrast, AGN have also been suggested to promote the SF in its host galaxy by enhancing the gas pressure in the ISM \citep[`positive feedback',][]{Silk:2013, Cresci:2015}.\\

The physical mechanisms through which AGN feedback acts are poorly constrained by observations. This includes how the energy is transported into the ISM through winds from the accretion disk, jets or the AGN radiation field, as well as how it couples to the surrounding ISM.
One important signature of AGN are powerful outflows which are a product from the energy released by the accreting SMBH and its interaction with the ISM. The outflows have a multi-phase nature.
In the X-ray, they can be identified in the form of fast radiation driven winds which originate from scales of BH accretion disc \citep{Tombesi:2010, Gofford:2013, Tombesi:2013}.
However, atomic, molecular, and ionised gas outflows are typically observed at distances from parsecs up to several kilo-parsecs from the galaxy centre. Ionised outflows are often identified from the systematically asymmetric emission line shape of [\ion{O}{iii}]$\lambda\lambda$4960,5007. The blue shoulder is usually interpreted as a bi-conical outflow of AGN-ionised gas where the receding side is obscured by dust from the host galaxy \citep[e.g.][]{Heckman:1981, Bischetti:2017, Bae&Woo:2016, Wylezalek:2020}.
Both observations \citep[]{Silk:1998, King:2003, Holt:2006, Holt:2008, Cicone:2014} and hydrodynamical simulations \citep[e.g.][]{DiMatteo:2005, Gaspari:2020} suggest that multi-phase outflows represent a key channel through which the AGN-feedback acts on the host galaxy. Moreover, such feedback can significantly stimulate turbulent condensation, leading to gas precipitation and chaotic cold accretion (CCA; e.g., \citealt{Gaspari:2019}).

Due to the compactness of the AGN-dominated region from which the outflows are launched, it is still a matter of debate how the evolution of the central BH impacts the host galaxy on different scales. 
An increasing number of studies provide a detailed and spatially resolved analysis of ionised gas outflows in local AGN \citep[e.g.][]{Greene:2011, Husemann:2016a, Revalski:2018, Venturi:2018, Husemann:2019, Riffel:2020}.
One big challenge is to determine the outflow location, orientation and intrinsic size.
Works from \cite{Greene:2011,Liu:2013, Harrison:2014, McElroy:2015} suggested that galaxy-scale outflows might be a prevalent in the majority of luminous AGN. In contrast, a recent study from \cite{Singha:2022} showed that among the luminous AGN in the representative sample of the Close AGN Reference Survey \citep[CARS,][]{Husemann:2022}, a large fraction (64\%) have ionised gas outflows with an extent less than $100 \,{\rm pc}$. This disagreement can partially be explained by beam-smearing in seeing-limited observations \citep{Husemann:2016b, Villar-Martin:2016, Davies:2019}.
Another problem is that the inferred outflow properties such as electron density, masses and kinetic energies are poorly constrained as they depend on the outflow geometry which is often unknown in luminous type 1 AGN \citep[e.g.][]{Rakshit:2018}.

Constraining the fundamental spatial, energy and timescales is crucial to understand how AGN-driven multi-phase outflows are launched and how they couple to the host galaxy ISM. 
For AGN in the Sloan Digital Sky Survey (SDSS) \cite{Mullaney:2013} and \cite{Woo:2016} have shown that outflows occur more frequently in AGN with a high Eddington ratio. 
In this regime, radiation pressure from the accretion disc couples to the dense ISM and can drive the outflow  \citep{Fabian:2012}, suggesting that ionised gas outflows are launched during the AGN phase. 
Among the general AGN population, narrow line \mbox{Seyfert 1} (NLS1) galaxies show low black hole masses at high accretion rates (e.g. \citealt{Boroson:2002,Grupe:2010}, review by \citealt[see review by][]{Komossa:2007}).
NLS1s often host ultra-fast outflows (UFOs) \citep[e.g.][]{Gupta:2013, Parker:2017, Reeves:2018, Kosec:2018, Reeves:2019, Xu:2022}.
Furthermore, NLS1s are located at the extreme end of the Eigenvector 1 (EV1) correlation which involves the widths and strengths of [\ion{O}{iii}]$\lambda 5007$, H$\beta$ and \ion{Fe}{ii} emission lines \citep{Boroson:1992}.
\cite{Wang:1996} extended the EV1 plane by the soft X-ray photon index where NLS1s typically exhibit a prominent soft X-ray excess \citep{Gliozzi:2020}.
The EV1 correlation is thought to be primarily driven by the Eddington ratio \citep{Sulentic:2000} which suggests that there might be a connection between BH accretion rate and the launching mechanism of a multi-phase outflow. 
For a comprehensive understanding, we need to resolve the multi-phase outflow both spatially and in its kinematic components.

In this work we present a detailed analysis of the ionised gas outflow in the centre of Mrk~1044.
Mrk~1044 is a nearby, luminous (\mbox{$L_{\rm{bol}} = 3.4\times 10^{44} \, \rm{erg\:s}^{-1} $}, \citealt{Husemann:2022}) NLS1 with a stellar mass of \mbox{$\log({M_\star / {\rm M}_\odot}) = 9.92^{+0.17}_{-0.12}$} \citep{Smirnova-Pinchukova:2022}. Its central engine is powered by a BH with a reverberation mapped mass of \mbox{$M_ \bullet=2.8\times10^6\,\rm{M}_\odot$} \citep{Du:2015}.
Mrk~1044 shows a soft X-ray excess which \cite{Mallick:2018} explained by relativistic reflections from a high density accretion disk.
From the several narrow absorption lines that are present in Mrk~1044's {\emph XMM-Newton} spectra, \cite{Krongold:2021} identified four kinematically distinct UFOs. Their velocity and density structure suggest that the two light absorbers may originate from the same multi-phase outflow.

The results presented in this work are based on the analysis of Mrk~1044's host galaxy presented in \citet[][hereafter Paper~I]{Winkel:2022} where we found that the host galaxy SF is concentrated in a circumnuclear ellipse (CNE) at $\sim 300 \,{\rm pc}$ from the centre.
Furthermore, we detected tentative signatures of ionised gas channelling towards the centre, which might be a signature of ongoing BH fuelling. Despite the high BH accretion rate, the host galaxy doesn't exhibit any signs of disturbance even at small distance from the nucleus ($\sim$ 160 pc).
Based on the velocity field of the stellar component we have constrained Mrk~1044's systemic velocity to $cz = 4913.4 \pm 0.2 \, {\rm km\: s}^{-1}$, corresponding to a redshift of $z=0.0164$. In the very centre of Mrk~1044, we detected a spatially resolved ionised gas outflow, emphasising the hypothesis of a complex multi-phase outflow.
In this work we use optical IFU data, UV spectroscopic data and radio imaging to constrain the properties of Mrk~1044's multi-phase outflow. Combining the observations we aim to get a detailed understanding of the outflow location, geometry and its physical driver. 
Throughout this paper we assume a flat $\Lambda$CDM cosmology with $H_0 = 70 \rm{kms}^{-1}\rm{Mpc}^{-1}$, $\Omega_M = 0.3$, and $\Omega_\Lambda=0.7$. In this framework, 1$\arcsec$ corresponds to 333$\,{\rm pc}$ at the galaxy redshift, where Mrk~1044's associated luminosity distance is 70.0$\,{\rm Mpc}$.

\section{Observations and data reduction}

\subsubsection*{Optical IFU observations}
\label{Sect:MUSE_NFM_observations}

We use integral field spectroscopic data obtained with narrow field mode (NFM) of the Multi Unit Spectroscopic Explorer \citep[MUSE,][]{Bacon:2010,Bacon:2014} at the Very Large Telescope (VLT). The data reduction is described in Paper~I.
The reduced data cube consists of $369 \times 378$ spaxels, corresponding to a FOV of $9\farcs23 \times 9\farcs45$. The spectral resolution is almost constant with a full width at half maximum (FWHM) of $2.54 \pm 0.10\,\textrm{\r{A}}$ across the entire wavelength range 4750$\,\textrm{\r{A}}$ - 9350$\,\textrm{\r{A}}$ corresponding to $160.4\,\rm{km\:s}^{-1}$ and $81.5\,\rm{km\:s}^{-1}$ at the blue and the red end of the spectrum, respectively. 
In Paper~I we describe the modelling of the adaptive optics shaped point spread function (PSF) using a hybrid approach between the empirical PSF and the analytical PSFAO19 model from \cite{Fetick:2019}.
The spatial resolution measured from the PSF extracted at the broad H$\beta$ emission is $89\,{\rm mas}$ which corresponds to $30\,{\rm pc}$ in the galaxy system.
We have deblended the AGN from the host emission in an iterative process that is described in Paper~I. In this work, we aim to extract the properties of the outflow from the original data cube that contains the blended emission from AGN and host.

\begin{figure*}
 \resizebox{\hsize}{!}{\includegraphics{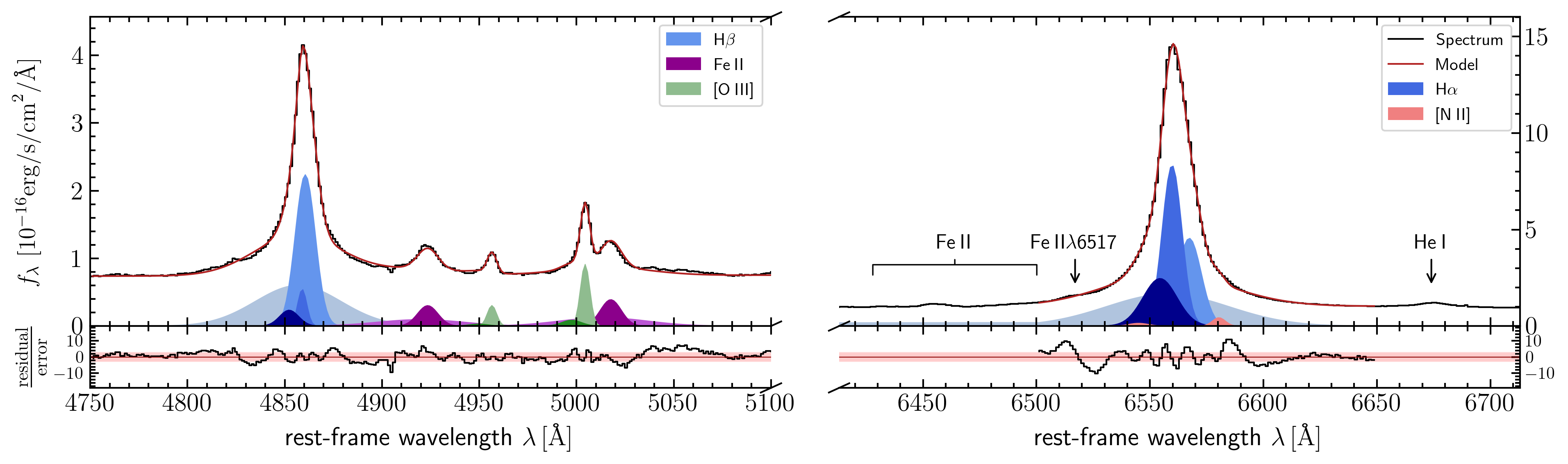}}
 \caption{Modelling Mrk~1044's AGN spectrum in the H$\beta$-[\ion{O}{iii}] region (left) and H$\alpha$-[\ion{N}{ii}] region (right). To reproduce the observed spectrum (black) we use a multi-Gaussian model for H$\beta$ (blue), [\ion{O}{iii}] (green), \ion{Fe}{ii} (purple), H$\alpha$ (blue, right panel) and [\ion{N}{ii}] (red). The best-fit spectrum is shown as a red line and well reproduces the prominent emission lines, including the blue shoulder of the [\ion{O}{iii}] narrow line.}
 \label{fig:MUSE_fit}
\end{figure*}

\begin{table*}
\caption{Best-fit parameters from modelling Mrk~1044's central highest signal-to-noise spectrum with multiple Gaussian components. The left columns list the emission lines in the H$\beta$-[\ion{O}{iii}] region, the right column the H$\alpha$+[\ion{N}{ii}] region. Velocities and dispersion are computed in the galaxy rest-frame. 
Lines that are listed together are kinematically coupled. Furthermore, we have kinematically tied the wing component between H$\beta$+[\ion{O}{iii}] and H$\alpha$. In each of the strong lines, the core component has a velocity that is significantly different from zero.}         
\label{tbl:AGN_fit}      
\centering                          
\begin{tabular}{c c c c c c c c c}        
&  &  & H$\beta$ window & 
& \hspace{5mm}
&  &  H$\alpha$ window & \\  
\hline             
component &\hspace{5mm}
& line & $v_r$ [$\rm km\:s ^{-1}$]  & $\sigma \,\, [\rm km\:s ^{-1}]$ 
& \hspace{5mm}
& line & $v_r$ [$\rm km\:s ^{-1}$]  & $\sigma \,\, [\rm km\:s ^{-1}]$ \\  
\hline                        
   broad &\hspace{5mm}&  H$\beta$+\ion{Fe}{ii} & -272 $\pm$ 11   &  1349 $\pm$ 14  
        &\hspace{5mm}&              H$\alpha$ & -232 $\pm$ 13 &  1232 $\pm$ 13
	\\
  medium &\hspace{5mm}& H$\beta$+\ion{Fe}{ii}  & -48 $\pm$ 4  &   281 $\pm$ 2  
        &\hspace{5mm}&              H$\alpha$ &    206 $\pm$   87   &   212 $\pm$ 48
	\\
    core &\hspace{5mm}& H$\beta$+[\ion{O}{iii}]  & -144  $\pm$ 5   &   165 $\pm$  17  
        &\hspace{5mm} & H$\alpha$+[\ion{N}{ii}]   & -140 $\pm$ 7  &   180 $\pm$ 12
	\\
    wing &\hspace{5mm}& H$\beta$+[\ion{O}{iii}]  & -560 $\pm$ 20  &  251 $\pm$ 31
        &\hspace{5mm}& H$\alpha$                & -560 $\pm$ 20  &   251 $\pm$ 31

          \\
\hline                                   
\end{tabular}
\end{table*}

\subsubsection*{UV spectroscopy}
\label{Sect:HST_STIS_observations}

We employ an archival UV spectrum of Mrk~1044 that was acquired with the Space Telescope Imaging Spectrograph (STIS) on the Hubble Space Telescope (HST). 
This data has first been presented in \cite{Fields:2005a} where the authors identified two outflowing absorbers with super-solar metallicity. 
Since we have a robust measure of Mrk~1044's systemic velocity from the stellar rotational field, we re-analyse the outflow kinematics. We retrieve the archival data from the Hubble Legacy Archive\footnote{\url{https://hla.stsci.edu/}} and stack the two data sets o8k4010-50 and o8k4010-60 taken with the G140M grating. 
The target was observed on UTC 2003 June 28 with the 52X0.2 aperture centred on Mrk~1044's nucleus. The two spectra were acquired with an exposure time of 1294$\,{\rm s}$ and 1440$\,{\rm s}$ respectively and have a resolution of $0.053\,\textrm{\r{A}}/{\rm px}$. The stacked spectrum extents from 1194.57$\,\textrm{\r{A}}$ to 1249.10${\,\textrm{\r{A}}}$ with a central wavelength of $1222\,\textrm{\r{A}}$.

\subsubsection*{Radio imaging}

Mrk~1044 has been observed with Karl G. Jansky Very Large Array (VLA) in the C-band (6 GHz) on October 30, 2016 and X-band (10 GHz) on January 15, 2017. The JVLA  observations were taken in the A configuration which has a maximum baseline of $36.4 \,{\rm km}$ and a minimum baseline of $0.68 \,{\rm km}$.
The A configuration observations are sensitive to emission on scales up to 9$\arcsec$ in C-band and up to 5$\farcs$3 at X-band. The three C-band scans have an integration time of $245\,{\rm s}$ each resulting in a total integration time $735\,{\rm s}$.
The two X-band scans have an integration time of $325 \,{\rm s}$ each. Together with two existing additional scans ($330\,{\rm s}$), the X-band has a total integration time of $1310 \,{\rm s}$.
3C48 and 3C138 were used as flux density calibrators. 
We reduced the JVLA data using the Common Astronomy Software Applications \citep[CASA,][]{McMullin:2007} pipeline version 6.2.1.7 and the dedicated CASA tools for VLA observations.

\section{Analysis and Results}

\subsection{Optical AGN spectrum}
\label{Sect:Optical_AGN_spec}

\begin{figure*}
 \resizebox{\hsize}{!}{\includegraphics{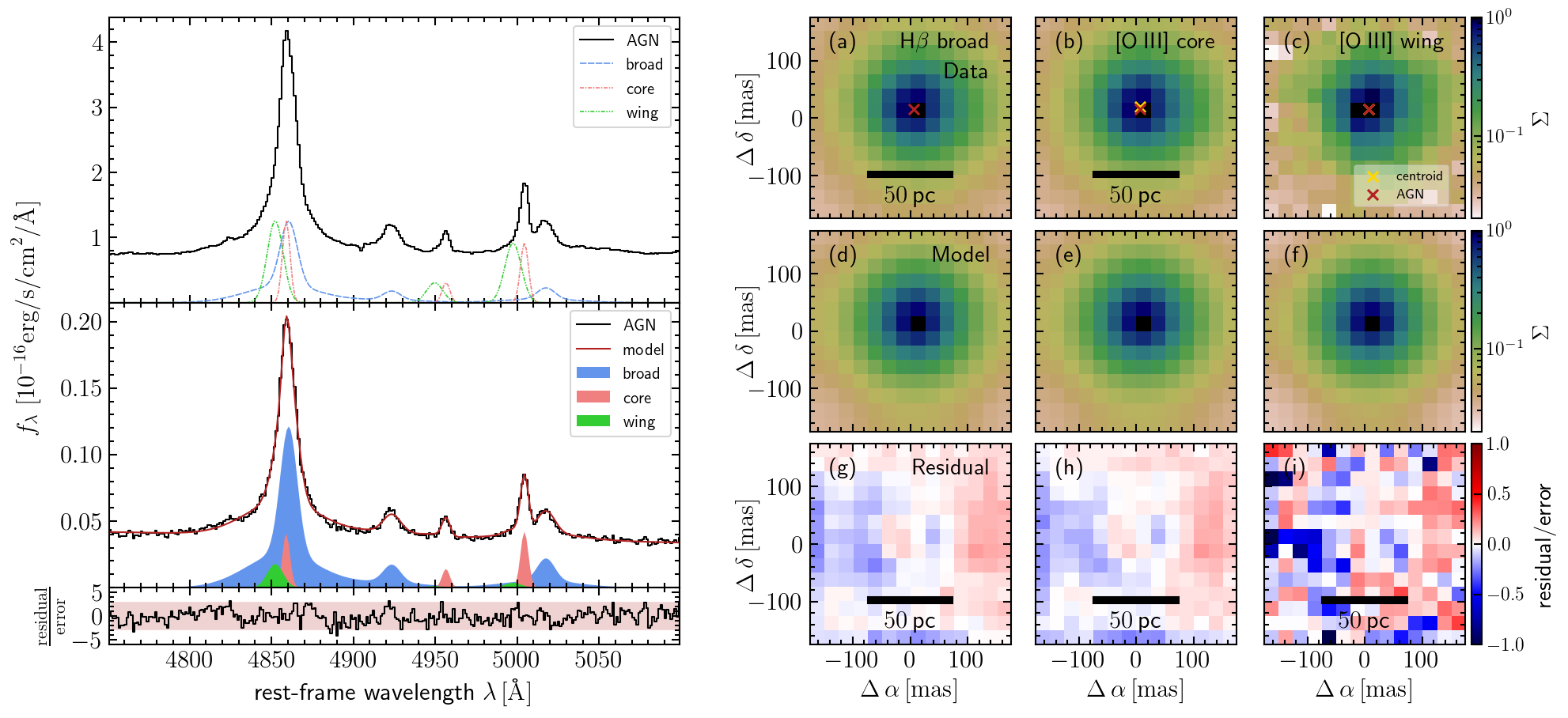}}
 \caption{Spectroastrometric analysis of Mrk~1044's central region using the original MUSE NFM-AO data cube. 
            The left panels show an arbitrary example spectrum that is picked from the small window around the nucleus shown in the panels on the right. To fit the spectrum, we keep the kinematics and line ratios of the kinematics fixed to that of the components found in the AGN spectrum.
            The panels on the right show the surface brightness within the central 150$\,{\rm mas}$ for the kinematic components where the flux maps are normalised to their peak flux.
            Here we only show the measure light distribution from the BLR H$\beta$ (left), [\ion{O}{iii}]-wing (middle), and narrow [\ion{O}{iii}]-core (right) component. From top to bottom the maps show the 2D light distributions, the best-fit model of the PSF and the residual maps. The red cross indicates the PSFAO19 centroid for the best-fit PSFAO19 fit to the BLR H$\beta$. The green and white crosses indicate the centroids for the PSFAO19 models to the surface brightness maps of the [\ion{O}{iii}]-wing and core respectively, in which the amplitude and position of the Moffat model were varied only. The offset of both [\ion{O}{iii}] core and wing component from the AGN position are smaller than $0.1\,{\rm px}$.}
 \label{fig:spectroastrometry_maps}
\end{figure*}

In order to understand the ISM properties of Mrk~1044's outflow on nuclear scales, we aim to extract the ionised gas kinematic components present in the innermost few pc. 
In Paper~I we have described the iterative deblending of AGN and host emission. For the following analysis we use the AGN spectrum from deblended AGN data cube which contains point-like emission from Mrk~1044's nucleus exclusively. Due to the finite width of the PSF, it probes the innermost $~30\,{\rm pc}$ around the nucleus. The spectrum in the observed frame is shown in Fig.~\ref{fig:MUSE_fit} for both the H$\beta$+[\ion{O}{iii}] and the H$\alpha$+[\ion{N}{ii}]+[\ion{S}{ii}] regions. It exhibits prominent emission lines H$\beta$/H$\alpha$ and [\ion{O}{iii}]$\lambda\lambda$4959,5007 and especially from \ion{Fe}{ii}, whereas the [\ion{N}{ii}]$\lambda\lambda$6548,6583 doublet is barely visible.

We simultaneously model the AGN spectrum in the H$\beta$+[\ion{O}{iii}] and the H$\alpha$+[\ion{N}{ii}] region. Our model consists of several Gaussian components and a linear approximation of the local continuum in each of the windows. To reproduce the H$\beta$ and  H$\alpha$ line shape, we require a broad+medium component and a narrow core component. The broad and medium component both have a kinematically tied component in the prominent broad \ion{Fe}{ii} $\lambda\lambda 4924,5018$ doublet. For the forbidden doublet emission lines [\ion{N}{ii}]$\lambda\lambda$6548,6583 and [\ion{O}{iii}]$\lambda\lambda$4959,5007 we couple the flux ratios to the theoretical prediction of 2.96 \citep{Storey:2000,Dimitrijevic:2007}. To reproduce the narrow [\ion{O}{iii}] line shape we require an additional blue-shifted 'wing' component which has a kinematically tied counterpart in H$\beta$ and H$\alpha$.
To find the best-fit parameters of the model we minimise the $\chi ^2$ of the residuals with a Levenberg-Marquardt algorithm. We estimate the uncertainties of the parameters by modelling 1000 artificial spectra generated from the 1$\sigma$ flux density errors in the AGN spectrum.
We note that the identification of the components in the emission line spectrum can also be reproduced with software packages that involve a \ion{Fe}{ii}-template to account for the strong \ion{Fe}{ii} contribution in the H$\beta$-[$\ion{O}{iii}$] window. However, the systematic uncertainty of the optional \ion{Fe}{ii} template provides a poorer description of the AGN spectrum than our multi-Gaussian model, especially in the [\ion{O}{iii}] region where we perform most of our diagnostics.
We note further that the results presented in the following do not change within the uncertainties if the emission from ions with different ionisation potentials are not kinematically coupled. In particular, the results do not change qualitatively if we release the kinematic coupling between the H$\beta$ and [\ion{O}{iii}] components and the H$\alpha$-wing and [\ion{N}{ii}] components respectively.

Fig.~\ref{fig:MUSE_fit} shows that the AGN spectrum in both spectral regions is well reproduced by our four component model. Considering the high signal-to-noise ratio (SNR), also the H$\alpha$ line shape is well-reproduced close to its centre. The broad wings, however, are contaminated by the broad \ion{Fe}{ii}$\lambda$6517, \ion{He}{i} emission and \ion{Fe} complex \citep[c.f. Fig.7][]{Veron-Cetty:2004} which are not included in our model. 
The kinematic parameters of the best-fit model of the AGN spectrum are summarised in Table~\ref{tbl:AGN_fit}. 
Both H$\alpha$+[\ion{N}{ii}] and H$\beta$+[\ion{O}{iii}] narrow core components have radial velocities significantly smaller than zero by 
\mbox{$ v_r({\rm H}\beta + [\ion{O}{iii}]) =-144  \pm 5 \, \rm km\:s ^{-1} $} 
and 
\mbox{$ v_r({\rm H}\alpha +[\ion{N}{ii}]) = -140  \pm 7 \, \rm km\:s ^{-1} $}, indicating that the region from which the narrow line emission emerges is moving towards the observer. 
In addition to the blue-shifted narrow core component, we detect a wing component in the H$\beta$ and [\ion{O}{iii}] emission with a velocity shift of $-560 \pm 20  \,\rm km\:s ^{-1}$. 
This feature is often recovered in AGN spectra and is typically interpreted as an ionised gas outflow moving towards the observer, where the receding side is obscured by dust around the nucleus \citep[e.g.][]{Heckman:1981, Boroson:2005, Zakamska:2003, Woo:2016}.

\subsection{Spectroastrometry}
\label{Sect:Spectroastrometry}

\begin{table}
\caption{Spectroastrometric measurement for the spatial offset between the kinematic components. By construction, the broad H$\beta$ and H$\alpha$ components are located at the AGN position.
}             
\label{tbl:spectroastrometry}      
\centering                          
\begin{tabular}{c c c}        
\hline             
component & offset [mas] &  offset [pc] \\    
\hline                        
   H$\beta$ broad &  0.00$\pm$0.54  &  0.00$\pm$0.19 \\
   {[\ion{O}{iii}]} core & 0.67$\pm$0.54  &  0.23$\pm$0.19 \\
   {[\ion{O}{iii}]} wing & 0.91$\pm$1.22  &  0.31$\pm$0.42
    \smallskip \\
    H$\alpha$ broad  &  0.00$\pm$0.05 & 0.00$\pm$0.02\\
    {[\ion{N}{ii}]} core  &  0.66$\pm$1.09 & 0.23$\pm$0.40\\
\hline                                   
\end{tabular}
\end{table}

\begin{figure*}
 \resizebox{\hsize}{!}{\includegraphics{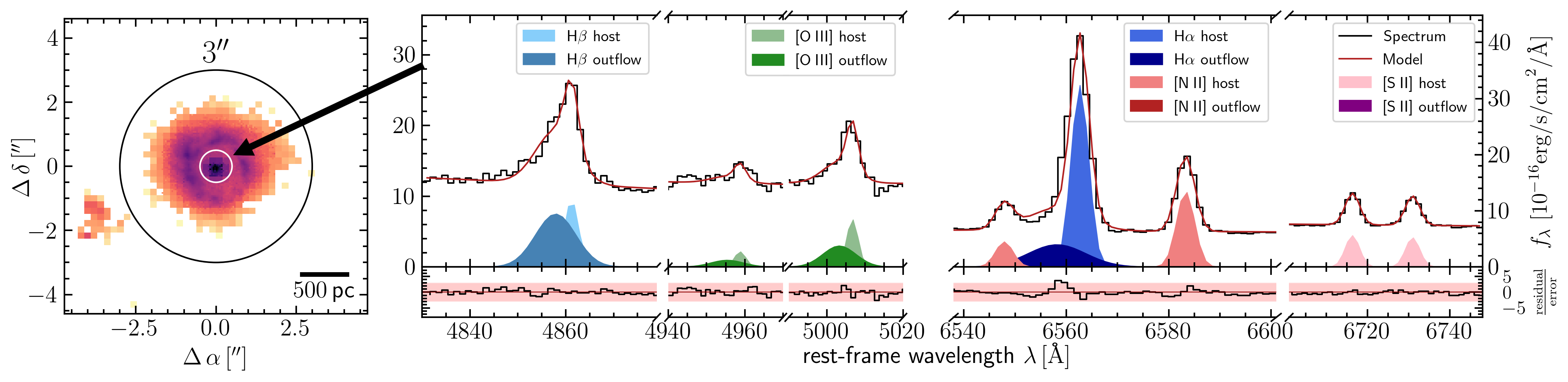}}
 \caption{Origin of the forbidden [\ion{S}{ii}] emission in Mrk~1044's center. The left panel shows the H$\alpha$ surface brightness map where the H0 outflow is located the central 0$\farcs$5. The right panel shows the two component model for the inner 0$\farcs$5 integrated spectrum where we fixed the host and outflow kinematics to what we found from the 3$\arcsec$ integrated spectrum. For the outflow, we do not find a significant contribution of [\ion{S}{ii}] and [\ion{N}{ii}] suggesting that the gas density lies above their critical densities.
 }
 \label{fig:nuclear_fit}
\end{figure*}

To trace the kinematic components with a high precision we employ a technique called spectroastrometry \citep[for discussion see][]{Bailey:1998,Gnerucci:2010,Gnerucci:2011}.

After having identified two outflowing systems in both Mrk~1044's H$\beta$- and [\ion{O}{iii}] emission, we now aim to constrain their location and spatial extent. 
For this purpose we have developed the {\large S}pectroastrometric analysis tool for {\large I}onized gas {\large E}mission in {\large N}earby {\large A}GN \citep[\texttt{Siena3D},][]{Siena3D}. \texttt{Siena3D} locates and traces kinematic components in 3D spectroscopic observations with high precision. The software package is available on GitHub\footnote{\texttt{\url{https://github.com/nicowinkel/Siena3D}}} under a OSI\footnote{\url{https://opensource.org/licenses}} approved licence.
The concept behind \texttt{Siena3D} is described in \cite{Singha:2022} where the technique has been applied and tested for the entire CARS sample of nearby unobscured AGN. Here, we give a brief description of the concept.

(1) As a first step, we use the kinematic components identified in the AGN spectrum to generate normalised base spectra (top left panel of Fig.~\ref{fig:spectroastrometry_maps}). All emission lines that belong to the kinematic components are combined in the respective base spectrum. Their flux ratios and kinematics are tied to what we measured in the highest signal-to-noise spaxel. This step involves the assumption that both line ratios and kinematics are constant throughout the structure that is traced (e.g. a compact fast moving ionised gas cloud or the BLR). 

(2) In the next step, we use the base spectra to fit the spaxels in a 14$\times$14 pixel (0$\farcs$35$\times$0$\farcs$35) aperture around the AGN position using a non-negative least squares optimiser. For one arbitrary spaxel, the best-fit model together with the contribution from the components are shown in the bottom left panel of Fig.~\ref{fig:spectroastrometry_maps}. Our results do not depend on the exact size of the aperture, since the high SNR core of the PSF (FWHM=$1.6...3.6\,{\rm px}$) is captured entirely.

(3) From the best-fit model we map the flux that originates from the individual kinematic components.
We fit the PSF to the light distributions to locate their centroids relative to the AGN position. For this task we adopt the hybrid PSF model that involves the empirical PSF and the analytic PSFAO19 model for the adaptive-optics shaped point-spread function measured in (in Paper~I, using the analytic model from \citealt{Fetick:2019}).

(4) Finally, we evaluate whether the light distribution deviates from a point-like emission to test whether the component has an intrinsic extended that is spatially resolved. The methodology is described in Appendix~\ref{appendix:size} where we constrain the projected extent of a resolved component (see Sect.~\ref{sect:resolved_outflow}). 

The precision by which different components can be kinematically and spatially disentangled depends on multiple parameters including the size and sampling of the PSF, the accuracy of the PSF model, the spectral resolution, the velocity offset between the kinematic components and the SNR of the emission lines and their dispersion.
In order to estimate the systematic uncertainty of our method specific to our data set, we have generated a set of mock data cubes by simulating the emission line profiles at different velocities, locations and sizes. 
We manually displaced the [\ion{O}{iii}]-wing component while keeping the location of the broad and core components fixed. 
At the SNR of [\ion{O}{iii}] core and wing component, the offset of both can be detected down to 0.1$\,{\rm px}$, corresponding to $2.5\,{\rm mas}$ or 0.85$\,{\rm pc}$ in the galaxy system.
Furthermore, we have simulated spatially resolved structure by convolving the PSF with a Gaussian kernel. For Mrk~1044 we can constrain the intrinsic extent of the outflow down to a size of 0.2$\,{\rm px}$ ($1.7\,{\rm pc}$).
At smaller scales, the precision is limited by the relatively low SNR of the wing component that is blended with the core component in each of the emission lines [\ion{O}{iii}]$\lambda\lambda$4959,5007.

For the Mrk~1044, the result of the spectroastrometric measurement in the H$\beta$-[\ion{O}{iii}]-region is shown in Fig.~\ref{fig:spectroastrometry_maps}, with the corresponding centroid positions listed in Table~\ref{tbl:spectroastrometry}. 
The broad component originates from the $\mu{\rm pc}$-scale BLR that is unresolved and follows the PSF light profile by construction. Therefore, the residual maps of the broad component reflect the systematic uncertainty of the PSF. But also the residuals of the [\ion{O}{iii}]-core and wing component show no extended structure. Neither [\ion{O}{iii}] outflowing component has an intrinsic extent that is larger than 0.2$\,{\rm px}$ (1.7$\,{\rm pc}$).
Furthermore, both core and wing component are located at the exact AGN position within the systematic uncertainties. We therefore conclude that neither of the [\ion{O}{iii}]-outflows have a projected distance from the nucleus that is larger than 0.85$\,{\rm pc}$.

\subsection{The spatially resolved outflow}
\label{sect:resolved_outflow}

During the analysis of the star-forming CNE in Paper~I, we have reported a high surface brightness spot in the centre that is present after subtracting the AGN emission. In the following we refer to this feature as H0 as it is most pronounced in H$\alpha$ and H$\beta$ surface brightness maps. The left panel of Fig.~\ref{fig:nuclear_fit} shows that in the host data cube the feature is constrained to the innermost $0\farcs$5.
In Paper~I the analysis with a single Gaussian emission line model yielded blue-shift of -140$\,{\rm km\:s}^{-1}$ and a peak velocity dispersion that is significantly higher than that of the CNE ($\sigma = 234 \pm 9 \,{\rm km\:s}^{-1}$, see fig.~7 in Paper~I). The high velocity-offset with respect to the galaxy systemic velocity suggests that the ionised gas is outflowing.
We detect the gas in each of the narrow emission lines H$\beta$, [\ion{O}{iii}], H$\alpha$ and  [\ion{N}{ii}] whereas we cannot trace a significant component [\ion{S}{ii}] given the low spectral signal-to-noise of the spatially resolved analysis. 

Although the outflow appears circularly symmetric, it is not an artefact from the PSF subtraction.
To better understand this we recall the PSF subtraction process.
H$\alpha$ and H$\beta$ are the wavelengths at which we have extracted the empirical PSF. By definition, the empirical PSF contains all the point-like emission. 
After the PSF cube construction, our PSF subtraction method is an iterative routine to clean the host emission from the AGN contamination.
Since we interpolate the PSF between H$\alpha$ and H$\beta$, an inaccurate PSF model would cause much stronger residual features at wavelengths at which the empirical PSF was not directly measured, that is [\ion{O}{iii}].
Furthermore, the central feature is present after each iteration of the PSF subtraction process. In particular, it is present after the first iteration where the host-contaminated AGN-spectrum is subtracted. 
The circularly symmetric surface brightness profile can thus be explained by a compact source together with the effect of beam-smearing.
To understand the origin of this spatially resolved outflow, we constrain its extent as described Appendix~\ref{appendix:size}. Our analysis yields a projected intrinsic extent of $0.55 \pm 0.10 \,{\rm px}$, corresponding to $4.6 \pm 0.6 \,{\rm pc}$ in the galaxy frame.

\subsubsection*{Multi-component analysis}
\label{sect:two_component_analysis}

\begin{figure*}
\centering
 \resizebox{.8\hsize}{!}{\includegraphics{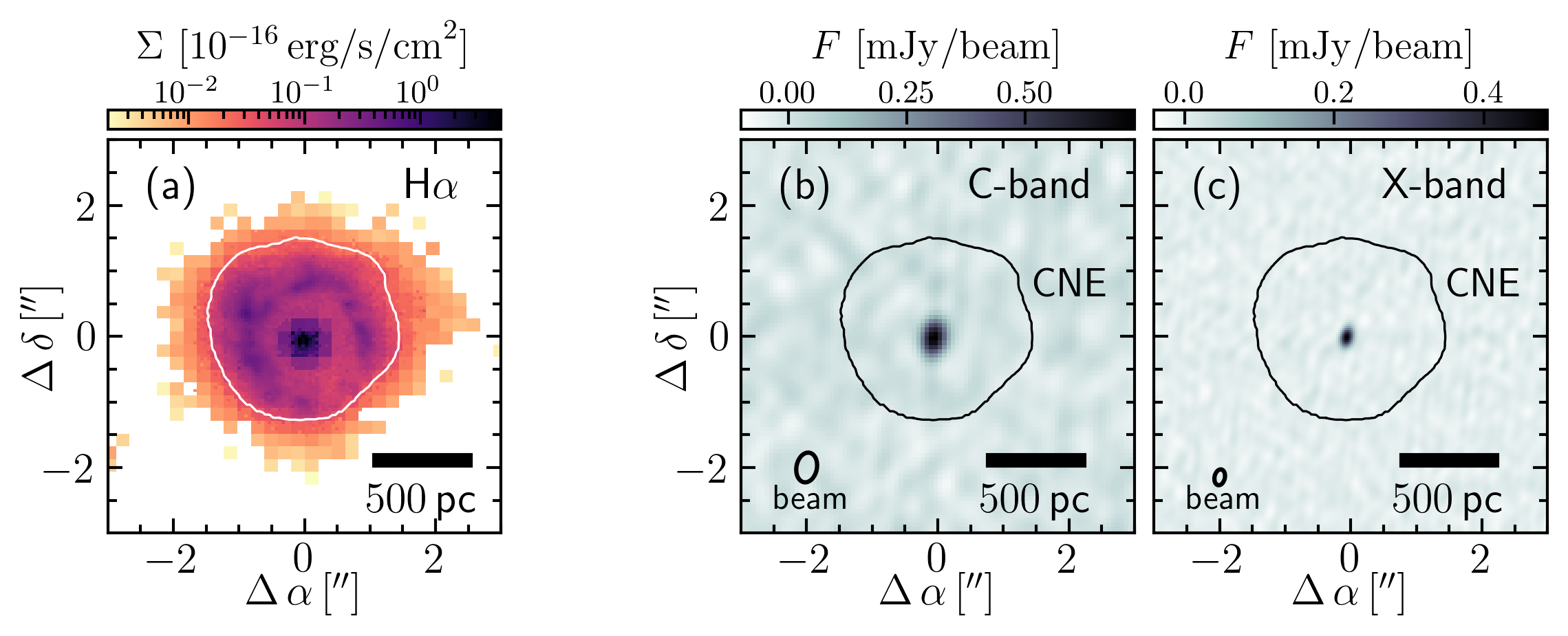}}
   \caption{Mrk~1044's nuclear region seen in the optical (left) and radio emission (right).
   Panel (a) shows the surface brightness map of the H$\alpha$ narrow line emission extracted as described in Paper~I. 
   The panels on the right show the broad-band EVLA images of the radio emission in the C-band (b) and X-band (c) with the corresponding FWHM-contour of the beam shown in the lower left corner respectively. The contours of the star forming CNE encompassing the nucleus are shown as black contours (see Paper~I). In both C- and X-band the nuclear radio emission is unresolved by VLA.}
 \label{fig:radio_images}
\end{figure*}

The spectrum around the nucleus is a superposition of the emission from the host and that of the spatially resolved outflow H0. The two signals stem from intrinsically distinct structures,  potentially located at different regions in the galaxy and having different ionisation conditions.
To disentangle and analyse their signal, we follow the methodology described in \cite{Singha:2022}, which contains the following steps:

(1) We fit H$\beta$ + $\ion{O}{iii} \lambda\lambda$4959,5007 with a two-component (host and outflow) Gaussian model each in a within a 3$\arcsec$ aperture. This gives us a robust estimate of the integrated host and outflow kinematic parameters.

(2) We fit each of the H$\beta$ + [$\ion{O}{iii}] \lambda\lambda$4959,5007 and \mbox{H$\alpha$ + [$\ion{N}{ii}] \lambda\lambda$6548,83 + [$\ion{S}{ii}] \lambda\lambda$6716,31} with the two-component model, this time keeping the kinematic parameters fixed to the values retrieved in step 1. Furthermore, we couple the line ratios among the [\ion{O}{iii}] and [\ion{N}{ii}] doublet lines to the theoretical prediction of 2.96 \citep{Storey:2000,Dimitrijevic:2007}.
Due to the relatively low SNR of the blended emission lines, this task could only be performed on the $0\farcs5$-integrated spectrum. 

The result of the process in shown in Fig.~\ref{fig:nuclear_fit}. Each of the strong emission lines is detected at a >3$\sigma$ confidence level with their kinematic parameters listed in Table~\ref{tbl:Outflow_summary}.
Except for the high signal-to-noise centre of the H$\alpha$ emission line, the residual spectrum is uniformly distributed around zero, with a scatter of 0.22$\sigma$ (FWHM), where $\sigma$ is the formal uncertainty from the AGN-blended cube. We conclude that the multi-Gaussian model provides a good description of the nuclear host spectrum. We detect a wing component that is present in H$\beta$, H$\alpha$ and [$\ion{O}{iii}$] exclusively. In contrast, the corresponding wing components in [$\ion{N}{ii}$] and [$\ion{S}{ii}$] are faint or missing such that we could only estimate upper flux limits.
We constrain and discuss the underlying ionisation mechanism from the emission line ratios in Sect.~\ref{Sect:nuclear_excitation_mechanism}.

\subsection{Radio emission}
\label{Sect:Radio_Emission}

The EVLA broad-band images in the C- and X-band are shown in Fig.~\ref{fig:radio_images}. The peak frequencies are located at 5$\,{\rm GHz}$ (C-band) and 10$\,{\rm GHz}$ (X-band). At these frequencies, we detect the AGN with a peak flux of \mbox{$758\pm19$}$\,\mu$Jy/beam (C-band) and \mbox{$496\pm10$}$\,\mu$Jy/beam (X-band).
The radio emission appears to be concentrated in the beam aperture in both bands. 
To quantify the size of the source, we use the CASA routine \texttt{imfit} to measure the AGN emission within a 10$\,{\rm px}$ radius from the centre which is sufficient to capture the source's entire emission. We find that the source is spatially unresolved by VLA and the centroid of its radio emission matches the the location of the AGN within the uncertainties.
Its integrated flux density amounts \mbox{$714\pm32$}$\,\mu$Jy in the C-band and \mbox{$484\pm18$}$\,\mu$Jy in the X-band respectively.
Assuming that the spectrum between 5$\,{\rm GHz}$ and 10$\, {\rm GHz}$ is described by a power law $S_{\nu} \propto \nu^{\alpha}$, we estimate the spectral index from the flux density. Its uncertainty depends both on the statistical error from thermal noise of the image and the systematic uncertainties $\sqrt{ \sigma_{\rm rms}^2 + \sigma_{\rm sys}^2 }$. Combining both, we use eq.~(1) from sect.~3.2 from \cite{Ramirez-Olivencia:2022} to estimate the uncertainty throughout our calculation. From Mrk~1044's integrated flux densities we obtain a spectral index of \mbox{$\alpha_{\rm int} = -0.56 \pm 0.23$}, whereas using the peak flux densities yields \mbox{$\alpha_{\rm peak} = -0.61 \pm 0.16$}.
Both values are consistent with each other, indicating once more that the source is spatially unresolved by VLA. As discussed by \cite{Ramirez-Olivencia:2022}, the peak flux is not affected by the systematic uncertainty from the different beam sizes between the C- and the X-band. We therefore consider the more robust result to be  $\alpha_{\rm peak}$ and adopt its value for the following analysis.

\subsection{Ly-$\alpha$ absorption}
\label{Sect:model_Lya_absorption}

\begin{figure}
 \resizebox{\hsize}{!}{\includegraphics{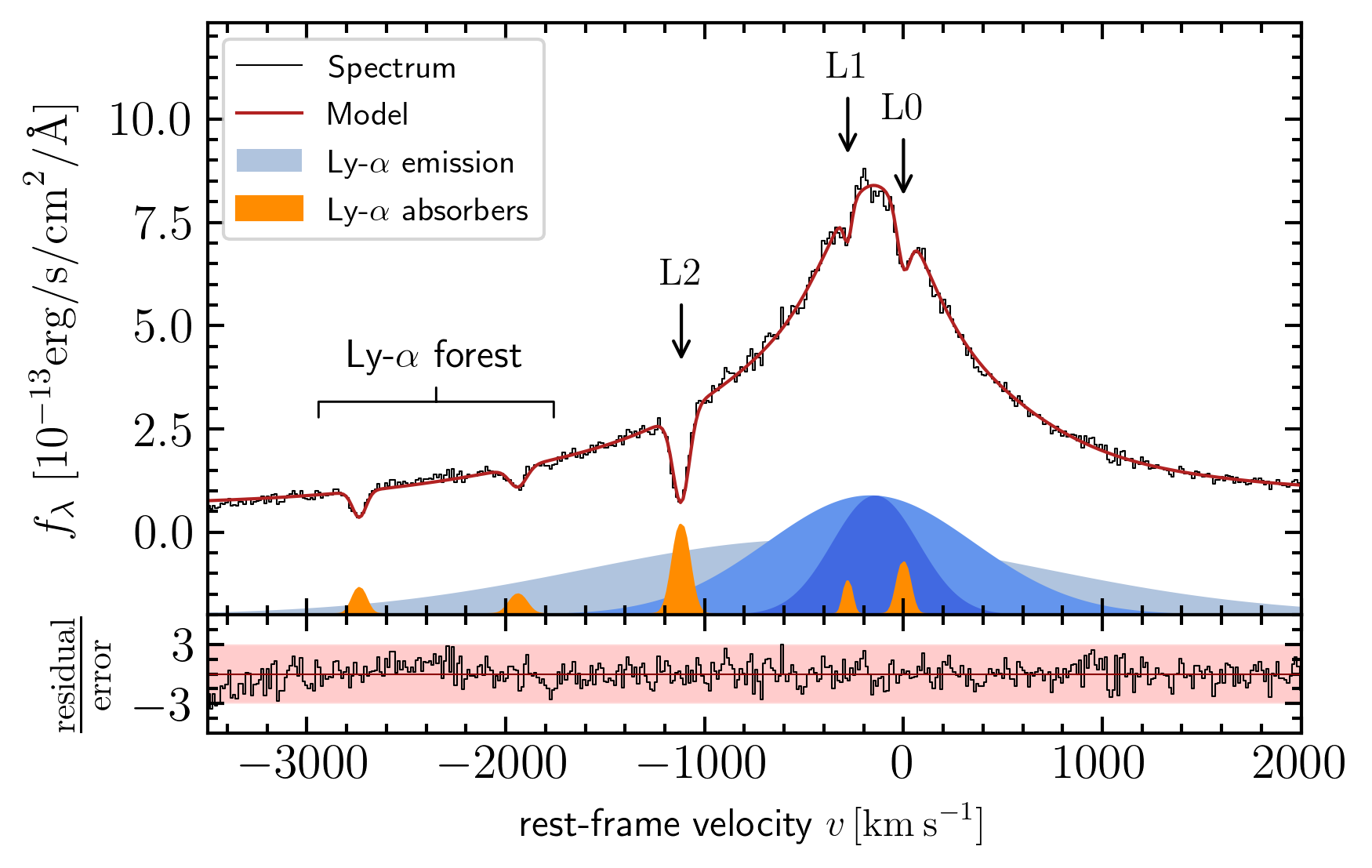}}
 \caption{Modelling Mrk~1044's UV spectrum in the Ly-$\alpha$ window. We use a multi-Gaussian model for the Ly-$\alpha$ emission (blue), and the absorbing systems (orange). The best-fit model (red line) reproduces the spectrum (black line) well in the selected wavelength range. Since the Ly-$\alpha$ systems L1 and L2 are accompanied by high metallicity absorption lines at the same velocity \citep{Fields:2005a}, the absorbing clouds must be local to the galaxy.}
 \label{fig:UV_fit}
\end{figure}

\begin{table*}
\caption{Summary of the outflowing systems. The only component that we spatially resolve are the H$\alpha$ and H$\beta$ core components in the AGN spectrum (Fig.~\ref{fig:MUSE_fit}). Their kinematic parameters listed here correspond to those of the core component of the two-component $0\arcsec\!5$ integrated two-component fit described in Sect.~\ref{Sect:nuclear_excitation_mechanism}. The [\ion{O}{iii}] emitters O1 and O2 are spatially unresolved. Their maximum offset and size equals the systematic uncertainty of our measurement. For the Ly-$\alpha$ absorbers, the spatial extent is not constrained.}             
\label{tbl:Outflow_summary}      
\centering                          
\begin{tabular}{c c c  c c c c c }        
\hline             
System&detection & location [pc] & extent [pc]  & $v_r$ [$\rm km\:s ^{-1}$]  & $\sigma \,\, [\rm km\:s ^{-1}]$
\\    
\hline                        
   H0 & H$\beta$, H$\alpha$,[\ion{O}{iii}] & <2.1 & $4.6 \pm 0.6 $ & $ -211 \pm 22$  &  $218 \pm 7$ \\
   O1 & H$\beta$, H$\alpha$,[\ion{O}{iii}] & <0.85 & <1.7 $\,{\rm}$ & $-144  \pm 5 $  &  $ 165  \pm  17$  \\
   O2 & H$\beta$, H$\alpha$,[\ion{O}{iii}] & <0.85 &  <1.7 $\,{\rm}$ & $-560 \pm 20 $ &  $251 \pm 31$
   \smallskip \\
   L1 & Ly-$\alpha$ & - & - & $-278 \pm 43$ &  $17 \pm 6$  \\
   L2 & Ly-$\alpha$ & - & - & $-1118 \pm 2$  & $41 \pm 5$ \\
\hline                                   
\end{tabular}
\end{table*}

\cite{Fields:2005a} have identified two outflowing systems in Mrk~1044's UV spectrum from Ly-$\alpha$, \ion{N}{iv}$\lambda$1239,43 and \ion{C}{iv}$\lambda$1549,51 absorption. Using the line strengths and a photo-ionisation model they estimated several times solar metallicities for both absorbing clouds and concluded that the absorbers were ejected from the BLR local to the galaxy.
In contrast to \cite{Fields:2005a} , we have used a kinematic model of a thin rotating disk to describe the stellar rotation in Paper~I. This represents a robust measure for the galaxy systemic velocity \mbox{$cz = 4913.4 \pm 0.2\, {\rm km\:s}^{-1}$} which allows for a self-consistent comparison between Mrk~1044's AGN, outflow and host galaxy properties after re-analysing the kinematics of the two UV absorbing systems.

We model Mrk~1044's UV spectrum in the Ly-$\alpha$-window using a set of Gaussians and a linear approximation of the continuum. To reproduce the broad emission line shape, we require three components whereas a single Gaussian is sufficient to model the absorption line-shapes. To find the best-fit parameters, we use the same technique that is described in Sect.~\ref{Sect:Optical_AGN_spec}. The result of this process is shown in Fig.~\ref{fig:UV_fit}. For the absorbers that we employ for the following discussion, the kinematic parameters for the individual components are listed in Table~\ref{tbl:Outflow_summary}. 
The broad component appears blue-shifted in the spectrum since Ly-$\alpha$ is a resonant line where the photons gain momentum by scattering on the fast-moving BLR clouds. Another reason might be that the BLR geometry might not be entirely symmetric. If the receding side of the BLR is obscured by dust, the observed Ly-$\alpha$ emission line is skewed towards the blue.

There are two absorbers with velocities $-1937 \pm 6 \,{\rm km\:s}^{-1}$ and $-2736 \pm 4 \,{\rm km\:s}^{-1}$ which can be explained by foreground extinction from the Ly-$\alpha$ forest \citep{Fields:2005a}.
The different systematic velocity explains the nature of the Ly-$\alpha$ absorbing system L0 that is located at the galaxy rest-frame with $v_r({\rm L0}) =  0.9 \pm 3.8 \,{\rm km\:s^{-1}}$.
Most importantly, we recover the two outflowing systems L1 and L2 for which we measure systematically higher velocities for the absorbing features compared to \cite{Fields:2005a}.
Both systems are local to the galaxy as they show corresponding absorption from \ion{N}{v} and \ion{C}{iv}. We measure the outflow velocity of the absorbing systems L1 and L2 as $v_r({\rm L1}) = -278 \pm 43 \, {\rm km\:s}^{-1}$ and $v_r({\rm L2}) = -1118 \pm 2 \, {\rm km\:s}^{-1}$, respectively.

\section{Discussion}

There are multiple outflowing systems present in Mrk~1044's outflow, both in optical emission and UV absorption. In Table~\ref{tbl:Outflow_summary} we summarise their kinematic properties and their distance from the nucleus. In the following we discuss the ionised gas excitation mechanism close to the galaxy nucleus and limitations of the diagnostics. Furthermore, we discuss whether we can link the different outflow velocities between [\ion{O}{iii}]-emitters and Ly-$\alpha$-absorbers using a geometric model. Finally, we estimate the mass outflow rates and energetics and discuss the future impact of the multi-phase outflow on the host galaxy star formation.

\subsection{Nuclear excitation mechanism}
\label{Sect:nuclear_excitation_mechanism}

In Paper~I we have shown that Mrk~1044 has a high star formation rate that is concentrated in a  circumnuclear ellipse (CNE). The 3$\arcsec$-integrated emission line ratios have shown that all spaxels in Mrk~1044's CNE are located in the star forming regime. 
But also in the very centre at the location of the outflow, the line ratios seem to favour SF as the dominant excitation mechanism.
However, after disentangling the host emission from the resolved outflow H0, a closer look onto the forbidden line ratios suggests that the situation is more complicated.

\subsubsection{Nuclear star formation hidden by the outflow}
\label{sect:nuclear_star_formation}
\begin{figure}
 \centering
 \resizebox{.7\hsize}{!}{\includegraphics{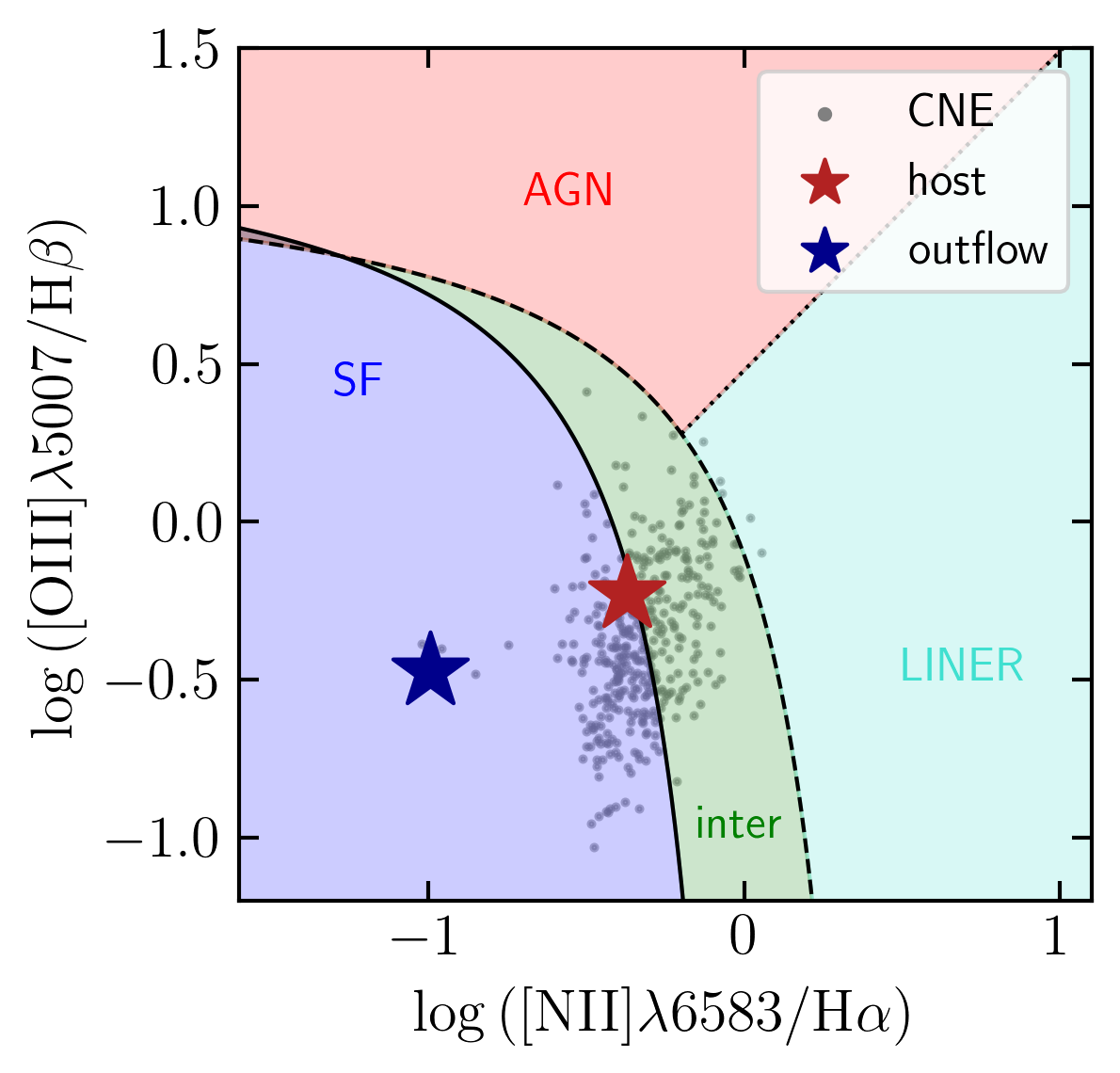}}
 \caption{BPT diagnostics of the extended outflow H0. The grey data points correspond to the 8$\times$8 co-added spaxels that belong to the star-forming CNE (see Paper~I). The stars represent the two components that are present in the $0\farcs5$-integrated spectrum that is shown in Fig.~\ref{fig:nuclear_fit}. The host component (red star) is consistent with SF-classification of the line ratios recovered in the CNE. For the outflowing component (blue star), both [\ion{O}{iii}] and [\ion{N}{ii}] emission might be suppressed due to collisional de-excitation in the high density gas. The BPT classification of the outflow as SF-ionised might therefore be misleading.
 }
 \label{fig:BPT}
\end{figure}

From the multi-component model of the central region in Sect.~\ref{sect:two_component_analysis}
we compile the BPT diagnostic diagram that is shown in Fig.~\ref{fig:BPT}. It contains the data from the spatially resolved 8$\times$8 co-added spaxels that belong to the star-forming CNE (see Paper~I) together with the 0$\farcs$5-integrated values for the host and outflow component.
By definition, the host component is located in the galaxy rest-frame. It occupies the same area as the resolved CNE-spaxels. Although the host component may contain a significant amount of beam-smeared emission from the CNE, the host [\ion{S}{ii}] emission peaks in the centre, which implies that it cannot be beam-smeared emission alone. Thus, even within the central 0$\farcs$5 there must be gas present at relatively low densities $n_{\rm e} < 4\times 10^3 \,{\rm cm}^{-3}$ the critical density of [\ion{S}{ii}] \citep{DeRobertis:1986}).

Combining the above findings, it indicates that the host galaxy star-formation in the CNE is not limited to the distance of $\sim 300 \,{\rm pc}$ from the nucleus.
It extends down to <0$\farcs$5 (<160$\,{\rm pc}$) from Mrk~1044's nucleus, but is heavily blended by the beam-smeared emission of the ionised gas outflow. Following the same procedure as in \cite{Winkel:2022}, we use the $0\farcs$5-integrated line flux from the host galaxy to estimate the ${\rm SFR} = 5 ^{+5}_{-3} \times 10^{-2} \,{\rm M}_\odot {\rm yr}^{-1}$ that is ‘hidden’ by the beam-smeared emission from the luminous ionised gas outflow.

\subsubsection{High outflow gas densities}
Compared to the host components, the line fluxes associated with the outflowing 'wing'-component H0 have a larger systematic uncertainty in the region around the narrow H$\alpha$+[\ion{N}{ii}]. This is caused by the blended emission from host and outflow components between [\ion{N}{ii}] and H$\alpha$ and the relatively low signal-to-noise. While we detect the [\ion{O}{iii}]-wing, the kinematically corresponding [\ion{N}{ii}] emission is weak such that we could only estimate an upper limit from at the level of the spectral noise (see Sect.~\ref{sect:two_component_analysis}).
The strong line ratios appear to unambiguously locate Mrk~1044's ionised gas outflow in the SF regime of the BPT diagnostic. 
We argue, however, that the BPT-classification of the outflow is misleading.

The total ionised gas mass ejected by the outflow from $< 4.6\,{\rm pc}$ is $~10^3\,{\rm M}_\odot {\rm yr}^{-1}$. Comparing with the relation for SF-driven outflows, the mass outflow rate exceeds the prediction by four orders of magnitude \citep[][see also Sect.~\ref{Sect:Outflow_energetics}]{Fluetsch:2019, Stuber:2021}. 
We conclude that Mrk~1044's outflow cannot be star-formation driven but must be powered by the radiation field of the luminous AGN. In this scenario, however, the AGN photo-ionisation is expected to dominate the emission line ratios.

A physical explanation that resolves the contradiction might be related to the compactness of the outflow. Considering that it has not yet expanded to host-galaxy scales beyond a few pc, the entire gas must be contained in a significantly smaller volume compared to the kpc-scale outflows of the extended narrow line region (ENLR). At less than a few ${\rm pc}$ distance from the nucleus we therefore expect significantly higher outflow gas densities.
If it exceeds the critical density of [\ion{N}{ii}] and [\ion{O}{iii}], the bulk of the ions collisionally de-excite and thus suppress the forbidden emission lines.
Assuming an electron temperature of $T_e=10^4 \,{\rm K}$ we estimate the electron density to be between the critical densities of [\ion{N}{ii}] and [\ion{O}{iii}] respectively, that is \mbox{$ 8.7\times 10^4 \,{\rm cm}^{-3}< n_{\rm e} < 7\times 10^5\, {\rm cm}^{-3}$} \citep{DeRobertis:1986}.
In this scenario the BPT classification of the outflow as SF-ionised can be explained by the high gas density, even if the actual excitation mechanism is related to the AGN radiation field.
Therefore, the BPT diagnostic is not a reliable tool to constrain the underlying ionisation mechanism in the presence of high density ionised gas dominating an unresolved structure.
Combining the above arguments we suggest that Mrk~1044's ionised gas outflow is likely AGN-driven.

\subsection{Exploring a geometric model for the unresolved outflow component}
\label{Sect:Geometric_Model}

In \ref{Sect:Optical_AGN_spec} and Sect.~\ref{Sect:model_Lya_absorption} we have described that in each [\ion{O}{iii}]-emission and Ly-$\alpha$ absorption, two spatially unresolved outflowing systems are present. A summary of the derived kinematic parameters is listed in Table~\ref{tbl:Outflow_summary}. 
We notice that velocity ratios among the [\ion{O}{iii}]-emitters ($v_{\rm O2}/v_{\rm O1} = 3.9$) and the Ly-$\alpha$ absorbers ($v_{\rm L2}/v_{\rm{L1}} = 4.0$) are similar. Furthermore, the velocity ratios between the faster and the slower components respectively match ($v_{\rm L2}/v_{\rm O2} = 2.0$, $v_{\rm L1}/v_{\rm O1} = 1.9$).  Motivated by this constant factor, we aim to identify the systems with each other and discuss whether the integrated [\ion{O}{iii}] emission line shape can be explained through a geometric alignment.

In order to constrain the geometry, we use a forward modelling approach to predict the shape of the [\ion{O}{iii}] emission lines based on an expanding shell geometry. 
We parameterise the model by the inclination $i$ of the ionised gas outflow velocity with respect to the line-of-sight, the 'narrowness' $\alpha / \theta_0$ that describes the radial dependence of the shells' intrinsic luminosity profile in units of the half-opening angle $\theta_0$, and the flux ratio between the two shells $r$.
We use a forward modelling approach to constrain the parameters by fitting the [\ion{O}{iii}] emission line shape. A detailed description of the method is described in the Appendix~\ref{appendix:shell_model}. 

The geometric model predicts highly inclined shell caps ($i=79\pm 8 ^{\circ}$) which are centrally concentrated $\alpha/\theta_0 = 0.2 \pm 0.1$.
Within the uncertainties, the result on the inclination is independent of the exact value of the BLR opening angle and how centrally concentrated the outflow is ($0.01 < \alpha/\theta_0 < 1$).

\subsubsection*{Expanding shell model versus dust torus model}

\begin{figure}
 \centering
 \resizebox{.8\hsize}{!}{\includegraphics{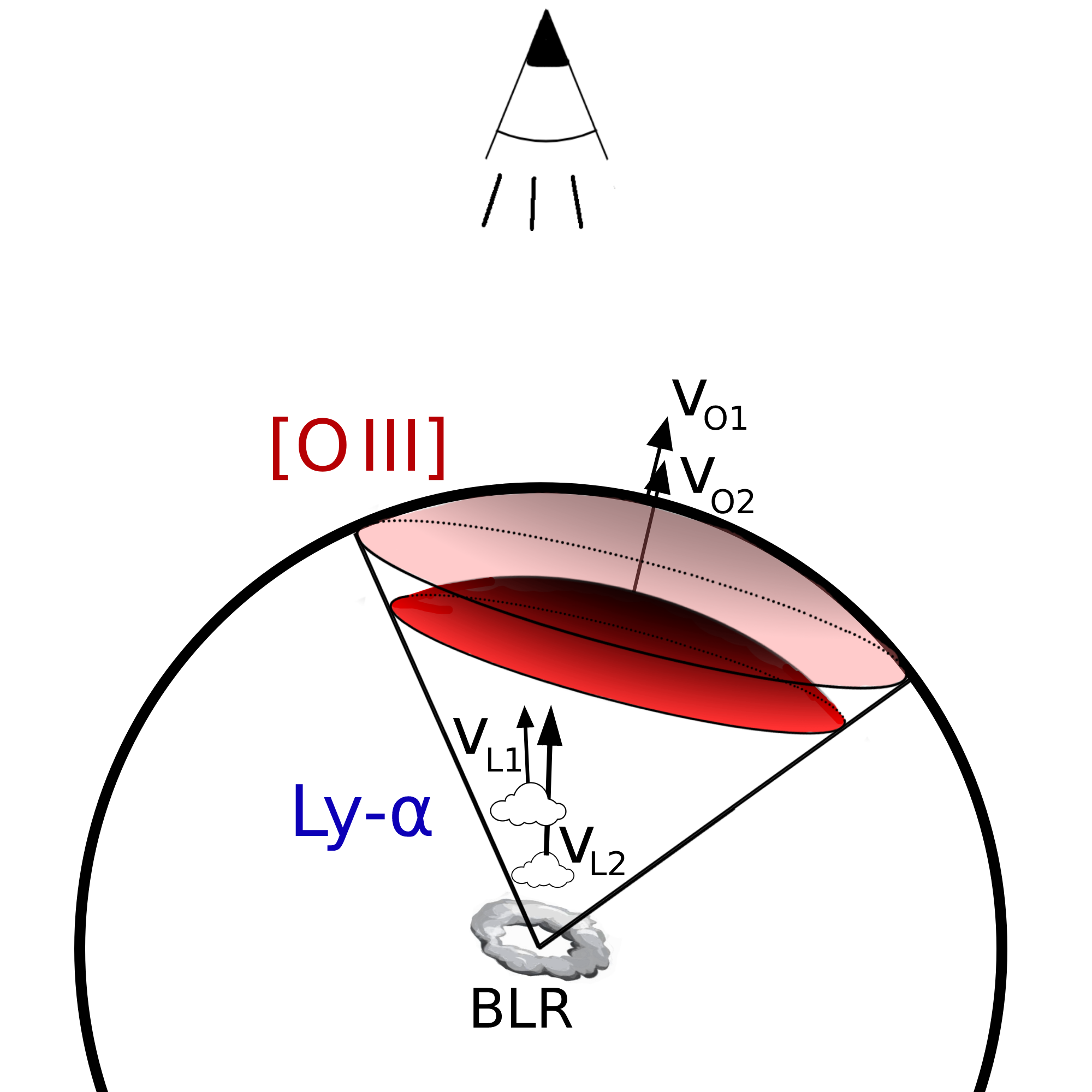}}
 \caption{Qualitative illustration of a possible alignment of the unresolved outflowing systems. Combining the dust torus model with the expanding shell model, the Ly-$\alpha$ absorption cannot be spatially linked with the [\ion{O}{iii}] emitting cloud. Hence, the inclination of the system as well as the geometry of the [\ion{O}{iii}] emitting structure is unconstrained. Compared to the extended [\ion{O}{iii}] emitting region, the Ly-$\alpha$ clouds are likely to be located closer to the AGN engine due to their higher velocity.}
 \label{fig:outflow_sketch}
\end{figure}

The AGN unification model predicts that an active nucleus possesses a compact strong radiation source around the SMBH that is surrounded by a high velocity emission-line region, the BLR. 
The relatively small scatter of the half-opening angles of the AGN dust-tori of $\sim 60 ^\circ$ \citep{Ricci:2017} suggests that the optical classification is a reliable indicator for the orientation of the obscuring dust torus.
For Mrk~1044's multi-phase outflow, the expanding shell model fails to reproduce the [\ion{O}{iii}] emission line shape for $i \lessapprox 60^{\circ}$.
At larger inclination, however, the dust-obscuration from the torus encompassing the nucleus is expected to block the view into the BLR which contradicts the type 1 nature of Mrk~1044.

One way to resolve the tension is that Mrk~1044's BLR obscuration may differ from the nearby low-luminosity Seyfert galaxies for which the AGN unification model was originally developed.
From the Swift-Burst Alert Telescope (BAT) selected local AGN sample, \cite{Ricci:2017} reported that the AGN obscured fraction is mainly determined by the Eddington ratio $\lambda_{\rm Edd}$. They concluded that the main physical driver of the torus diversity is the accretion rate, which regulates the torus covering factor by means of radiation pressure.
A clumpy torus \citep{Schartmann:2008, Stalevski:2019} provides a more accurate description of the relationship between X-ray and mid-IR luminosities \citep{Nenkova:2008,Stalevski:2012, Honig:2017} and also been resolved for individual systems with high-resolution mid-IR imaging \cite[e.g.][]{Isbell:2021}. It is unclear whether the same orientation-based unification model involving a smooth torus is directly applicable to a highly accreting system like Mrk~1044.
Nonetheless, \cite{Marin:2016} argued that the probability to see a type-1 source at high inclination is almost zero. Based on the polarisation properties of AGN in the 3CRR catalogue the authors concluded that the optical classification is a reliable indicator for the BLR orientation.

A less drastic solution for the mismatch between the inclination and Mrk~1044's unobscured nucleus is to discard our toy model's rigorous assumptions on the geometry multi-phase outflow. 
Since we have not spatially resolved any of the components, the Ly-$\alpha$ absorbing clouds may not be linked to the [\ion{O}{iii}] emitting clouds.
The different kinematics between the Ly-$\alpha$ absorbers and [\ion{O}{iii}] emitters could thus be a reflection of the inhomogeneity of the ISM on sub-pc scales regarding the ionisation conditions and the density distribution.
Compared to the [\ion{O}{iii}] ionised gas, the Ly-$\alpha$ absorbers have higher velocity but smaller dispersion. This can be understood if the outflow is launched from accretion disk scales and entrains the ambient gas, thus gradually decreasing its bulk velocity while increasing the dispersion/turbulence due to the interactions with the surrounding ISM \citep{Gaspari:2017}. This can also potentially trigger a CCA condensation rain (see Section~\ref{sect:expansion_to_galaxy_scales}).
In an alternative interpretation, the outflow is intrinsically elongated but aligned with the line-of-sight. Due to the optically thin [\ion{O}{iii}] clouds, a velocity gradient within the medium would increase the integrated observed line width and mimic a high dispersion of the spatially unresolved structure.
Both explanations favour that the [\ion{O}{iii}] emission occurs at larger distance from the nucleus than the Ly-$\alpha$ absorption. 
A qualitative illustration of this alignment is shown in Fig.~\ref{fig:outflow_sketch} where the fast Ly-$\alpha$ absorbing clouds are located closer to the nucleus. We note that the inclination and geometry of the [\ion{O}{iii}] clouds remain unconstrained. For the following discussion, we therefore do not adopt the specific value that predicted by the expanding shell model.

\subsection{Outflow energetics and mass outflow rates}
\label{Sect:Outflow_energetics}
 
 \begin{table}
\caption{Outflow energetics and outflow rates. From left to right the columns describe the quantity, the unit and the derived value for the ionised gas outflow system H0, O1 and O2 respectively. We compute the mass loading factor for the scale over which the resolved outflow is beam-smeared as well as for the integrated star forming CNE ($\eta_{\rm CNE}$) and the entire host galaxy ($\eta_{\rm tot}$) which have not yet been reached by the expanding outflow.}             
\label{tbl:energetics}      
\centering                          
\begin{tabular}{c c c c c c c c}        
\hline             
   &  & H0 &  O1   &  O2
\\    
\hline                        
   $A_{{\rm H}\alpha}$ & [mag] & 1.5 & 3.7 &  2.2 \\
   ${\rm log}(L_{{\rm H}\alpha})$ & $[{\rm erg\:s}^{-1}]$ &
    39.3 & 42.4 &   41.5 \\ 
   $M_{\rm ion}$ & $[10^5\,{\rm M}_\odot]$ & 0.01 &  10.5 &   1.2 \\
   $\dot{M}$ & $[{\rm M}_\odot \,{\rm yr}^{-1}]$& 0.002 & 15  &   7 \\
   ${\rm log}(\dot{p})$ & [${\rm dyne}$] & 30.5 &  34.1  &  34.4 \\
   ${\rm log}(\dot{E})$ & [${\rm erg\:s}^{-1}$] & 37.6 &  41.0   &   41.8 \smallskip \\
   $\eta_{0\farcs5}$ & & 0.05 &  276   &   127 \\
   $\eta_{\rm CNE}$ & & 0.01 &  80   &   37 \\
   $\eta_{\rm tot}$ & & 0.005 &  21   &   10 \\
   
\hline                                   
\end{tabular}
\end{table}

Despite the compactness of the ionised gas outflows at present day, it will propagate through the ISM of the host galaxy.
Once the shock front reaches the molecular gas from which the CNE produces stars, the outflow's energy and momentum will induce turbulence into the ISM (e.g., \citealt{Wittor:2020}).
To estimate the future impact on the host galaxy star formation, we compute the current mass, energy and momentum ejection rates released through the ionised gas outflow.
We stress that the following derivations require an assumption about the morphology of the [\ion{O}{iii}] emitting clouds and their velocity as derived in Sect.~\ref{Sect:Geometric_Model}. 
As discussed in the previous section, the Ly-$\alpha$ absorbers may not be spatially associated with the ionised gas outflow.
We therefore use the velocities of the [\ion{O}{iii}]-emitters for the following derivations.

As a first step, we derive the total  H$\alpha$-flux of the outflowing components O1 and O2 by integrating over the individual components' flux maps extracted in Sect.~\ref{Sect:Spectroastrometry}.
From the Balmer-decrement H$\alpha$/H$\beta$ and the Milky Way-like attenuation curve \cite{Cardelli:1989}, we derive the attenuation analogously to Paper~I.
As described in \cite{Husemann:2016a}, the ionised gas mass can then be estimated from the extinction-corrected intrinsic luminosity $L_{{\rm H}\alpha}$ as 

\begin{align}
    M_{\rm ion} &= 3.4 \times 10^6 \left( \frac{100 \, \rm{cm}^{-3}}{n_{\rm e}} \right) \left( \frac{L_{{\rm H}\alpha}}{10^{41} \rm{erg\:s}^{-1}} \right) {\rm M}_\odot .
\end{align}

Due to the missing [\ion{S}{ii}] and [\ion{N}{ii}] emission (see Sect.~\ref{Sect:nuclear_excitation_mechanism}), we assume a lower limit on the electron density \mbox{$n_{\rm e} = 8.7\times 10^4 \,{\rm cm}^{-3}$}, the critical density of [\ion{N}{ii}]. In this way we ensure that the outflow masses and energies are not underestimated.

The estimated mass outflow rate depends on the assumed outflow morphology \citep{Cicone:2014,Veilleux:2017}. Since the UV absorbers have relatively low velocity dispersion, we consider thin shells to be the most probable description of the [\ion{O}{iii}] outflow morphology.
However, in order to achieve a consistent comparison between our results and the scaling relations from \cite{Fiore:2017}, we assume a conical outflow geometry which lowers the derived values by a factor of approximately three.
We compute the ionised gas mass outflow rate as
\begin{align}
\dot{M}_{\rm OF,cone} = 3 \times \left( \frac{v_{\rm OF}}{100 \, {\rm km\:s ^{-1}}} \right) 
\left( \frac{M_{\rm ion}}{10^7 \,{\rm M}_\odot}\right) 
\left( \frac{1\,{\rm kpc}}{R_{\rm OF}}\right)
{\rm M}_\odot {\rm yr}^{-1}
\end{align}
where $v_{\rm OF}$ is the outflow velocity and $R_{\rm OF}$ is the size of the homogeneously filled cone.

In addition to the mass outflow rate, we estimate the momentum injection rate  $\dot{p}_{\rm OF} = v_{\rm OF} \dot{M}_{\rm OF}$ and the  kinetic energy injection rate as  $\dot{E}_{\rm kin} = 0.5 v_{\rm OF}^2 \dot{M}_{\rm OF}$.
Finally, we compare the SFR on different spatial scales and the mass injection rate from the AGN-driven outflow by computing the mass loading factor as $\eta = \dot{M}_{\rm OF}/ {\rm SFR}$.

The results for each of the outflowing components are listed in Table~\ref{tbl:energetics}. 
The energetics of Mrk~1044's outflow properties are affected by several systematic uncertainties. 
Especially the electron density $n_{\rm e}$ and the size of the system might underestimate each of the derived quantities, as we could only estimate lower and higher boundaries respectively. 
Nonetheless, the derived outflow properties hold some important implications especially for the future impact on the host galaxy discussed in Sect.~\ref{sect:expansion_to_galaxy_scales}.
The two unresolved [\ion{O}{iii}] are luminous and dominate the outflow mass and energy budget by four orders of magnitude. The total kinetic energy of the outflows is $ 7.3\times10^{41} \, \rm{erg\:s}^{-1}$ which corresponds to 0.2\% of Mrk~1044's bolometric luminosity ($ 3.4\times10^{44} \, \rm{erg\:s}^{-1}$, \citealt{Husemann:2022}). This favours the possibility of an energy-driven ionised gas outflow, if the photons couple with a comparable efficiency to the gas in the host galaxy ISM.

\subsection{Origin of Mrk~1044's radio emission}
Due to the radio-quite nature Mrk~1044 and the lack of extended radio jets, its nuclear radio emission could be produced by a multitude of processes. This includes the accretion disc coronal activity, an AGN driven wind, free-free emission from photo-ionised gas, or nuclear star formation. 
The radio spectra of \ion{H}{ii} regions is shaped by the free-free emission which, in the optically thin limit, produces a nearly flat radio spectrum with $\alpha \sim -0.1$. Together with the steep synchrotron spectrum from supernova remnants, the integrated radio emission of \ion{H}{ii} regions is expected to have a characteristic spectral index of $\sim -0.7$ which is consistent with the observations of star-forming galaxies \citep{Condon:1992, Panessa:2019, Perez-Torres:2021}.
To test whether the unresolved nuclear SFR is sufficient to produce the Mrk~1044's nuclear radio emission, we estimate the SFR from the observed radio continuum luminosity and the calibration from \cite{Murphy:2011}

\begin{equation}
\begin{split}
\frac{\rm SFR}{{\rm M}_\odot {\rm yr}^{-1}} 
= &\left[ 2.18 \left( \frac{T}{10^4 \,{\rm K}} \right)^{0.45} \left( \frac{\nu}{\rm GHz} \right)^{-0.1}
 + 15.1 \left( \frac{\nu}{\rm GHz} \right)^{\alpha^{\rm NT}}  \right]^{-1} \\
 & \times \frac{L_\nu}{ 10^{27}{\rm  erg\: s}^{-1}\:{\rm Hz}^{-1}}.
\end{split}
\label{eq:Murphy2011}
\end{equation}

Here, $\alpha^{\rm NT}$ corresponds to the non-thermal spectral index for which we use the value $\alpha_{\rm peak} = -0.61 \pm16$ measured between 5$\,{\rm GHz}$ and 10$\,{\rm GHz}$ (see Sect.~\ref{Sect:Radio_Emission}).
Using Mrk~1044's C-band luminosity and an electron temperature of $10^4\,{\rm K}$ we use this relation to estimate a star formation rate of \mbox{${\rm SFR} = (0.81 \pm 0.45) \times 10^{-3}\,{\rm M}_\odot {\rm yr}^{-1}$}. 
The associated uncertainty involves both the error of our measured parameters (see Sect.\ref{Sect:Radio_Emission}) together with the 50 percent uncertainty from relation (\ref{eq:Murphy2011}).
Since the SFR estimated from the radio luminosity is exceeded by the 0$\farcs$5 unresolved SFR derived from Mrk~1044's H$\alpha$-luminosity by one order of magnitude (see Sect.~\ref{sect:nuclear_star_formation}), we cannot exclude that the radio emission is partially produced by SF-related processes.
Thus, we conclude that neither the spectral index nor the amount of flux allows us to distinguish whether Mrk~1044's radio emission originates exclusively from AGN-related processes or the nuclear star formation.

\subsection{Outflow expansion from nuclear to galaxy scales}
\label{sect:expansion_to_galaxy_scales}

To test whether the individual components of the multi-phase outflow have the potential to reach galactic scales, we use the `k-plot' diagnostic from \cite{Gaspari:2018}. The k-plot compares the LOS velocity dispersion with the LOS velocity shift.
It is useful to assess the role of chaotic/turbulent motions and laminar/bulk motions, and related key physical processes such as chaotic cold accretion (CCA) and AGN outflows. We use the pencil-beam approach, which better traces individual clouds or small-scale gas elements (e.g., \citealt{Rose:2019,Maccagni:2021,North:2021,Temi:2022}).

\begin{figure}
 \centering
 \resizebox{.8\hsize}{!}{\includegraphics{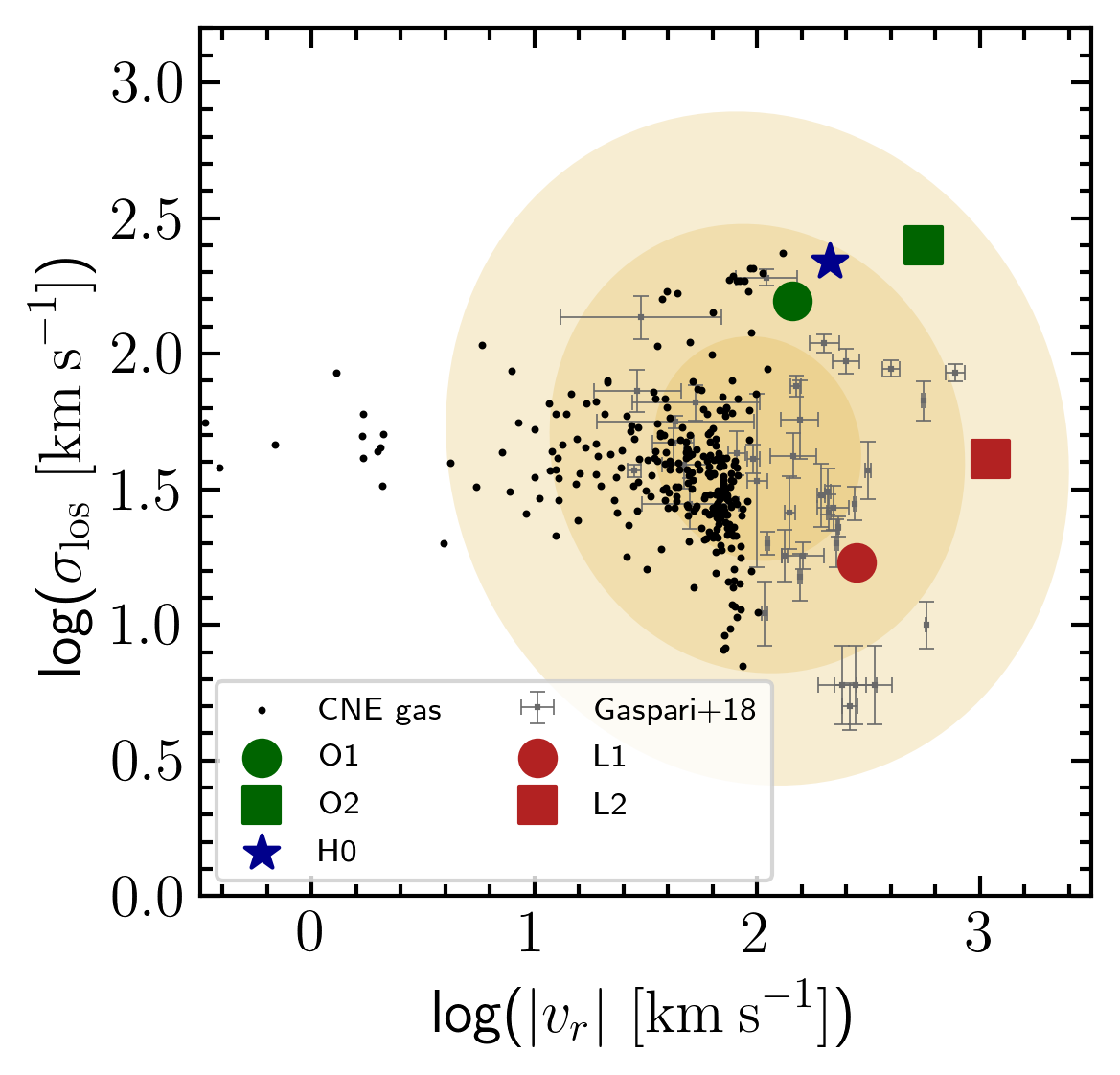}}
 \caption{Kinematical plot (k-plot) of the line broadening versus the line shift. It highlights the relative importance of chaotic/turbulent motions versus laminar/bulk motions and the connected physical processes.
 The yellow contours show the 1-3$\sigma$ confidence intervals for the chaotic cold accretion (CCA) feeding mode found by high-resolution hydrodynamical simulations \citep{Gaspari:2018}.
 The grey data points indicate measurements of warm and cold condensed gas in diverse galaxies (\citealt{Gaspari:2018}).  
 The ionised gas of Mrk~1044's CNE is shown as black data points. The tail towards lower velocity is mainly driven by an inclined rotating disk, with several points overlapping with the main CCA area. 
 In contrast, the four nuclear components (red and green points) have substantially higher bulk velocity indicating that they are not kinematically associated with the disk, but rather outflowing, in particular O2 and L2 (squares). O1 and L1 appear to interact more significantly with the surrounding medium, thus being more prone to turbulent condensation.
 }
 \label{fig:k-plot}
\end{figure}

Fig.~\ref{fig:k-plot} shows how the CNE gas (black points) and nuclear components (red/green) of Mrk~1044 compare with the multi-phase k-plot retrieved via high-resolution hydrodynamical simulations (yellow 1-3$\sigma$ contours; see \citealt{Gaspari:2018}) and related observational sample of diverse galaxies (grey).
The ionised gas in Mrk~1044 star-forming CNE with the highest velocity shows a symmetric spread of the dispersion within the 2$\sigma$ contours. 
The tail towards low velocities at constant dispersion is more heavily associated with the rotating disk. Despite the dominance of rotation, several black points reside well within the 1-$\sigma$ CCA region, suggesting that such gas elements can also be prone to turbulent condensation. This is a common scenario in spiral galaxies where the turbulent Taylor number is on average above unity, hence favouring a rotating CCA (\citealt{Gaspari:2015}).

Compared to the region the resolved data points of the rotating CNE occupy, the unresolved nuclear components separate.
The faster components \mbox{-- O2 and L2 --} are located in the 3$\sigma$ area and are thus expected to be less affected by condensation and CCA. This suggests that their momentum is currently little dissipated by the  interaction with the surrounding medium, which enables the outflows to escape at least from the high-density nuclear region. Due to their lower velocities, systems O1 and L1 are located within the 1.5$\sigma$ yellow confidence region. This can be interpreted that such outflowing components have started to deposit their kinetic energy into the ISM, increasing the turbulence of the ambient medium and potentially stimulating an initial CCA rain.

It is interesting to note that the [\ion{O}{iii}] outflows have systematically higher velocity dispersion than the rest of the system.
This can be interpreted in a few ways. On the one hand, the highly ionised outflow component could simply have a stronger intrinsic turbulence along its spine, which is indicative of the entrainment of surrounding material that increases the turbulence in the ionised gas. On the other hand, it could be interpreted as a projection effect if the [\ion{O}{iii}] gas velocity is directed at a small inclination with respect to the line-of sight. In this case the [\ion{O}{iii}] emission could be affected by beam
smearing as opposed to the kinematics from Ly-$\alpha$ foreground absorption. The observed [\ion{O}{iii}] line width might therefore reflect a systemic velocity gradient within the outflow and might not be interpreted as a dispersion.

\subsubsection{Future impact on the host galaxy}

\begin{figure*}
\centering
 \resizebox{\hsize}{!}{\includegraphics{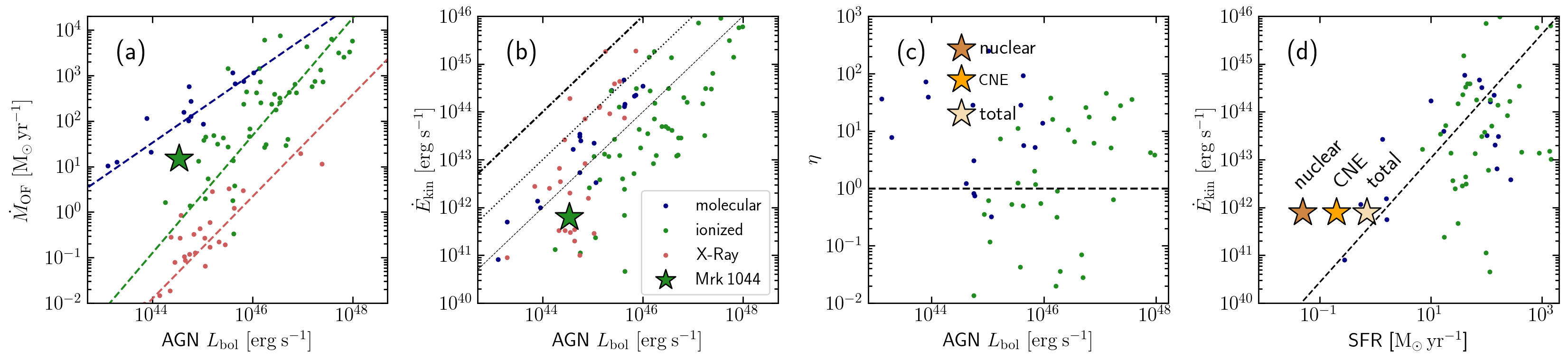}}
   \caption{Comparing Mrk~1044's integrated ionised gas outflow with AGN multi-phase outflow scaling relations. The coloured data points are taken from \cite{Fiore:2017} and show the molecular gas outflows traced by CO, OH (blue), ionised gas outflows (green) and X-ray outflows (red). 
   Panel (a) shows the outflowing ionised gas mass. The dashed coloured lines correspond to the correlations for the respective gas phase reported by \cite{Fiore:2017}.
   Panel (b) shows the AGN wind kinetic power where the dashed, dotted and dash-dotted line correspond to the $\dot{E}_{\rm kin} = 0.01,0.1,1 \, L_{\rm bol}$.
   Panel (c) shows the mass loading factor $\eta = \dot{M}_{\rm OF}/{\rm SFR}$. Here, the brown star represents Mrk~1044's SF in the immediate vicinity ($< 160\,{\rm pc}$) from the centre, the orange point to that of the SF in the CNE and the yellow point to that of Mrk~1044's galaxy-wide SFR.
   Panel (d) shows the kinetic outflow power as a function of SFR. 
   The solid red line indicates the predicted SFR from the AGN bolometric luminosity using the relation from \cite{Netzer:2009} together with the average $\dot{E}_{\rm kin}/L_{\rm bol} = 0.025$ for molecular outflows \citep{Fiore:2017}.
   Compared to the ionised gas outflows in the sample, Mrk~1044's outflow carries a lot of mass in the ionised gas phase. Its mass injection into the ISM exceeds the current SFR in the immediate vicinity of the nucleus, but also compared to the galaxy-wide values.
   }
 \label{fig:outflow_energetics}
\end{figure*}

Mrk1044's outflow carries a lot of mass in the ionised gas phase. To estimate its future impact on the host galaxy, we compare the summed energetics with the scaling relations of resolved galaxy-scale outflows.
The left panel of Fig.~\ref{fig:outflow_energetics} shows that compared to other AGN at similar luminosity, Mrk~1044's ionised gas outflow lies within the scatter of the outflowing mass - AGN bolometric luminosity correlation from \cite{Fiore:2017}.
It is more massive by $\sim 1 \,{\rm dex}$ than average, whereas the kinetic power is consistent with the ionised gas outflows in lower luminosity AGN. To compare the ionised gas outflow momentum rate with the AGN radiation momentum rate $L_{\rm bol}/c$, we compute the wind momentum load as $\dot{p}_{\rm OF}/\dot{p}_{\rm AGN} = 3.3$. It is consistent with the wide spread of the extended ionised gas outflows reported in \cite{Fiore:2017} and close to the expectation of $\dot{p}_{\rm OF}/\dot{p}_{\rm AGN} = 1$ for a momentum conserving outflow predicted by the \cite{King:2003} model.
Fig.~\ref{fig:outflow_energetics} highlights that multi-phase outflows often carry a large amount of their mass and kinetic power in the molecular gas phase, especially at low AGN luminosities \citep[see also][]{Fluetsch:2019,Veilleux:2020}. We therefore expect that the AGN-driven mass injection into the host galaxy ISM is even higher if Mrk~1044's outflow is accompanied by a molecular gas outflow.

An important difference when comparing Mrk~1044 with ionised gas outflows in nearby galaxies is that they extend over different spatial scales within the host galaxy.
While the relation from \cite{Fiore:2017} was calibrated for ionised gas winds that extend over $> 1\,{\rm kpc}$, Mrk~1044's ionised gas outflow is compact ($< 4.6\,{\rm pc}$) and located close to the galaxy nucleus ($< 2.1\,{\rm pc}$).
Its proximity to Mrk~1044's central engine is reflected in the mass loading factor $\eta = \dot{M}_{\rm OF} / {\rm SFR}$ which relates the SFR with the mass outflow rate. The intrinsic extent over which the galaxy property that is extracted, in this case the SFR, does affect the result dramatically, as it is shown in the third panel of Fig.~\ref{fig:outflow_energetics}. While the outflow is located at parsec-scale distance from the nucleus, the SFR can be measured in different apertures. 
If we compare the mass outflow rate with the SF in its immediate vicinity, that is the SF ‘hidden’ by the beam-smeared outflow (\mbox{$< 160\,{\rm pc}$}, see Sect.~\ref{Sect:nuclear_excitation_mechanism}), $\eta_{0\farcs5} = 276$ suggests that the outflow currently dominates the ISM properties of the host galaxy on a few $\sim 100\,{\rm pc}$ scales.
Such a high mass-loading factor is in-line with the picture of a young outflow that has recently been launched by the powerful AGN.
Furthermore, the ionised gas outflowing mass exceeds Mrk~1044's present day SFR, even if we compare it with the galaxy-wide SFR. However, this comparison should be regarded with caution as both quantities trace different regions of the host galaxy and the evolution of the outflow's properties during its propagation through the ISM are largely unknown. 

Since Mrk~1044's ionised gas outflow inclination and the morphology are unconstrained, it is not clear how the outflow will impact the host galaxy.
To estimate the maximum future impact on the host galaxy, we assume that the outflow is launched at large inclination relative to the galaxy rotation axis and carries its present-day mass and energy out to a few 100 pc from the nucleus. In this case it directly impacts the CNE and the galaxy disk where the host galaxy star formation is concentrated.
The total mass outflow rate also exceeds the integrated star formation rate in the CNE \mbox{($0.19 \pm 0.05 \,{\rm M}_\odot {\rm yr}^{-1}$)} and the galaxy-wide SFR \mbox{($0.70 \pm 0.17 \,{\rm M}_\odot {\rm yr}^{-1}$)} by more than one order of magnitude.
Mrk~1044's multi-phase outflow may therefore have the potential to deprive the host galaxy from their molecular gas reservoir and eventually quench the host galaxy star formation. 
Based on the mass loading factor \cite{Fiore:2017} argued that only powerful AGN are able to drive the co-evolution between SMBHs and galaxies. In this interpretation Mrk~1044 would represent an uncommon example where the host galaxy evolution is regulated by the central SMBH despite the relatively low luminosity of its AGN if compared to the overall AGN population.

In the other extreme scenario the outflow is orientated perpendicular to the disk, the cold gas by which the SF in the CNE is fuelled will not be affected by the energy injection. Hence, we expect that in this case the outflow's impact on the host galaxy is minor to none. The two extreme cases of outflow inclinations leave a wide range of possible scenarios in which the host galaxy SF may be quenched by the injection of turbulence into the ISM.
Resolving the inclination of the accretion disk relative to the galaxy plane is therefore crucial to estimate the future impact of Mrk~1044's outflow on the host galaxy.

\subsection{Mrk~1044 - a young AGN}
\label{sect:A_young_AGN}

In Paper~I we have touched on the idea that Mrk~1044 as a NLS1 could be archetypal for a population of ‘young’ AGN that host a rapidly growing SMBH. 
There we have shown that the circumnuclear SF enriches the ISM and shows potential signatures of ongoing BH feeding. SF is present even in the immediate vicinity of  Mrk~1044's nucleus (see Sect.~\ref{Sect:nuclear_excitation_mechanism}), and might play a role in channelling the gas towards the centre as suggested by hydrodynamic simulations \citep[][]{Davies:2007,Volonteri:2015,Zhuang:2020}. Since circumnuclear SF continues to enrich the ISM and has not yet been affected by the energy injection from the expanding multi-phase outflow, this process might continue until the outflow propagates through the ISM.

To estimate a frame for the look back time at which of the ionised gas outflow was launched, we assume a constant LOS velocity by which the outflow escapes from the central engine. Since the outflow inclination cannot be robustly constrained by the geometric toy model discussed in Sect.~\ref{Sect:Geometric_Model}, we assume that the outflow is launched perpendicular to the accretion disk. Further, we assume that the rotation axis of the accretion disk is aligned with that of the dust torus. In this case, the accretion disk inclination $i$ of a type~1 AGN cannot be larger than the half-opening angle of the dust torus, since the torus would obscure the view to the BLR.
We therefore assume Mrk~1044's outflow inclination to be smaller than $58^\circ$, the median of the dust torus half-opening angle distribution \citep{Ricci:2017}.
From the intrinsic velocity $v = v_{\rm LOS}/\cos i$ and the projected distance from the nucleus $d = d_{\rm proj}/\sin i$ we estimate the timescale as

\begin{align}
    t \lessapprox \frac{d_{\rm proj}}{v_{\rm LOS}} \tan i
\end{align}

With the maximum projected offset from the nucleus of 0.85$\,{\rm pc}$ (see Sect.~\ref{Sect:Spectroastrometry}) we infer that the two ionised gas outflow components were launched no longer than 9,900$\,{\rm yr}$ (O1) and 3,500$\,{\rm yr}$ (O2) ago.
Using the upper limit for the location of the spatially resolved outflow H0, we estimate an upper limit for its age of $\sim$25,000$\,{\rm yr}$.
The derived 'age' estimates should be regarded as an upper limit since interaction of the fast-moving ionised gas outflow with the host galaxy ISM would slow down the outflow during its expansion.
Furthermore, if the outflow velocity is directed closer to the observer (i.e. at smaller inclination $i$), the resulting timescale would be shorter.
The picture of Mrk~1044 as a young AGN is consistent with the undisturbed kinematics that we could trace down to $0\farcs5$ ($\sim 160\,{\rm pc}$) from the centre (see Paper~I).
Furthermore, the ongoing star formation in the CNE around Mrk~1044's nucleus (\mbox{$\sim 300\,{\rm pc}$}) has not been quenched by the energy injection from the multi-phase outflow.
Even within the innermost $160\,{\rm pc}$ where the ionised gas outflow dominates, the integrated line ratios that belong to the host are consistent with ionization through SF (see Sect.~\ref{Sect:nuclear_excitation_mechanism}).

Since the distance of an AGN-ionised gas cloud to the AGN engine directly translates into a light travel time, the ENLR size can be interpreted as a proxy for the lifetime of a single AGN episode \citep{Lintott:2009, Keel:2017, Husemann:2022}.
At fixed luminosity, AGN with more massive BHs would be statistically observed at much later times in their episodic phase than lighter BHs. This interpretation implies an evolutionary sequence for AGN where systems hosting less extended ENLRs have a shorter AGN lifetime or are ‘younger’.
The EV1 relation could be explained as a reflection of the time-dependence at which the AGN phase is observed. NLS1s, which are located in the tail of the EV1 relation, would represent the extreme end of the rest-frame difference between [\ion{O}{iii}]-core and systemtic redshift. 
The interpretation of NLS1s as ‘young’ AGN would explain the relative weakness of their narrow [\ion{O}{iii}] emission lines by a combination of {\it i)} the absence of an AGN-ionised ENLR in the host galaxy {\it ii)} the compactness of the ionised gas outflow and {\it iii)} the high gas densities in the compact outflow (see Sect.~\ref{Sect:nuclear_excitation_mechanism}).
In this scenario, the prevalence of strong broad \ion{Fe}{ii} emission in the spectra of NLS1s which can be interpreted as the finite timescale on which AGN quenching is able to shut off the host galaxy star formation on nuclear scales.

Another aspect of the interpretation is that it explains Mrk~1044's offset of the narrow [\ion{O}{iii}]-core from the galaxy systemic velocity. As we have demonstrated, the [\ion{O}{iii}]-core component is outflowing with $- 140\,{\rm km\:s}^{-1}$ (Sect.~\ref{Sect:Optical_AGN_spec}) which is fundamentally different from the many of AGN where the narrow [\ion{O}{iii}] traces the ENLR associated the galaxy rest-frame \citep[e.g.][]{Harrison:2016,Husemann:2022}.
If a young AGN drives the expanding outflow, this mismatch can be understood since the [\ion{O}{iii}]-core component traces a young structure that has a smaller intrinsic extent. Hence, the integrated narrow [\ion{O}{iii}] may trace different structures, depending on the age of the system.
This implies that the outflow velocities of ‘young’ AGN cannot be estimated from the integrated emission line spectrum alone. Spatially resolving young AGN host galaxies and their compact outflows is crucial to correctly interpret their spatial \emph{and} kinematic structure.

A coherent picture for the recent activity of Mrk~1044's black hole growth emerges. Combining the arguments of {\it i}) a compact outflow, {\it ii}) unperturbed host kinematics and {\it iii}) SF-dominated radiation field in the close to the nucleus, we conclude that Mrk~1044's AGN phase has been triggered recently. Following the idea that NLS1s host rapidly growing SMBHs at an early stage of their evolution, we predict that highly accreting AGN host galaxies with low-mass SMBHs show similar signatures close to the nucleus.

 \section{Summary and conclusions}

In this work we have combined IFU data from MUSE NFM-AO with the UV spectrum from HST to constrain the ionised gas properties of Mrk~1044's outflow. We have detected two outflowing systems in both ionised gas emission and Ly-$\alpha$ absorption that are unresolved at the spatial resolution limit of MUSE NFM-AO. We have used a spectroastrometric approach to locate the ionised gas outflows traced by [\ion{O}{iii}] emission. Our key results are the following.

\begin{itemize}

\item{Mrk~1044's narrow core component O1 that we detect H$\beta$, H$\alpha$ and [\ion{O}{iii}] emission is outflowing with \mbox{$v_{\rm O1} = -144 \pm 5 \,{\rm km\:s}^{-1}$}. The [\ion{O}{iii}]-wing component O2 represents an additional component with \mbox{$v_{\rm O2} = -560 \pm 20 \,{\rm km\:s}^{-1}$}.
}

\item{Both ionised gas outflows O1 and O2 are spatially unresolved and have a projected offset from the AGN location smaller than 1$\,{\rm pc}$. These two outflows carry the bulk of the ionised gas mass and energy and have been launched no longer than $\sim$10$^4\,{\rm yr}$ ago.
}

\item{Despite the compactness of Mrk~1044's ionised gas outflow, its mass outflow rate and kinetic energy injection rate are comparable to typical AGN-driven ionised gas winds at similar AGN luminosity. The high present-day outflow velocity and high mass loading suggest that the outflow has the potential to expand to galaxy scales and to impact the ISM properties.}

\item{Using a geometric toy model, the two Ly-$\alpha$ absorbers at velocities $-278 \,{\rm km\:s}^{-1}$ and $-1118 \,{\rm km\:s}^{-1}$ could not be kinematically associated with the [\ion{O}{iii}] emission. This stresses that the complex nature of the multi-phase ISM on sub-pc scales.}

\item{The gas densities in Mrk~1044's spatially resolved outflow H0 are above the critical density of the forbidden [\ion{S}{ii}] and [\ion{N}{ii}] line doublets, which implies that the classical BPT diagnostics cannot be used to constrain the underlying ionisation mechanism.}

\end{itemize}

Our results stress the importance of studying the AGN-host galaxy connection on different energy and density scales with spatially resolved observations. Although the ionised gas outflow can be identified from the AGN spectrum alone, Mrk~1044's compact multi-phase outflow stands out from well-studied examples of kpc-scale extended outflows. The different outflow components, kinematics and energetics highlight the complexity of the AGN feedback cycle, especially on the smallest galactic scales from where the outflow is launched.
To better understand the role of the warm ionised gas in the immediate vicinity of accretion-mode AGN, more IFU observations with high spatial resolution are required. Performing a similar analysis on a larger sample will help to constrain the impact of young, compact outflows on the host galaxy as well as the underlying physical mechanisms behind AGN feeding and feedback.

\begin{acknowledgements}
    We thank the anonymous referee for suggestions and comments that helped improve the presentation of this work.
    This work was funded by the DFG under grant HU 1777/3-1. 
    NW, BH and ISP acknowledge travel support from the DAAD under grant 57509925 and the hospitality of MS at the University of Manitoba where large parts of this paper were written. 
    BH was partially supported by the DFG grant GE625/17-1. 
    VNB gratefully acknowledges assistance from the National Science Foundation (NSF) Research at Undergraduate Institutions (RUI) grant AST-1909297. Note that findings and conclusions do not necessarily represent views of the NSF. 
    CO and MS are supported by the Natural Sciences and Engineering Research Council (NSERC) of Canada. 
    MG acknowledges partial support by NASA Chandra GO9-20114X and HST GO-15890.020/023-A, and the {\it BlackHoleWeather} program. 
    MPT acknowledges financial support from the State Agency for Research of the Spanish MCIU through the ''Center of Excellence Severo Ochoa'' award to the Instituto de Astrof\'isica de Andaluc\'ia (SEV-2017-0709) and through the grant PID2020-117404GB-C21 funded by MCI/AEI/10.13039/501100011033, by ERDF ''A way of making Europe''.
      
\end{acknowledgements}

%
%

\bibliographystyle{aa}
\bibliography{references.bib}

\begin{appendix}

\section{Size of the resolved outflow H0}
\label{appendix:size}

\begin{figure}
 \centering
   \includegraphics[width=\columnwidth]{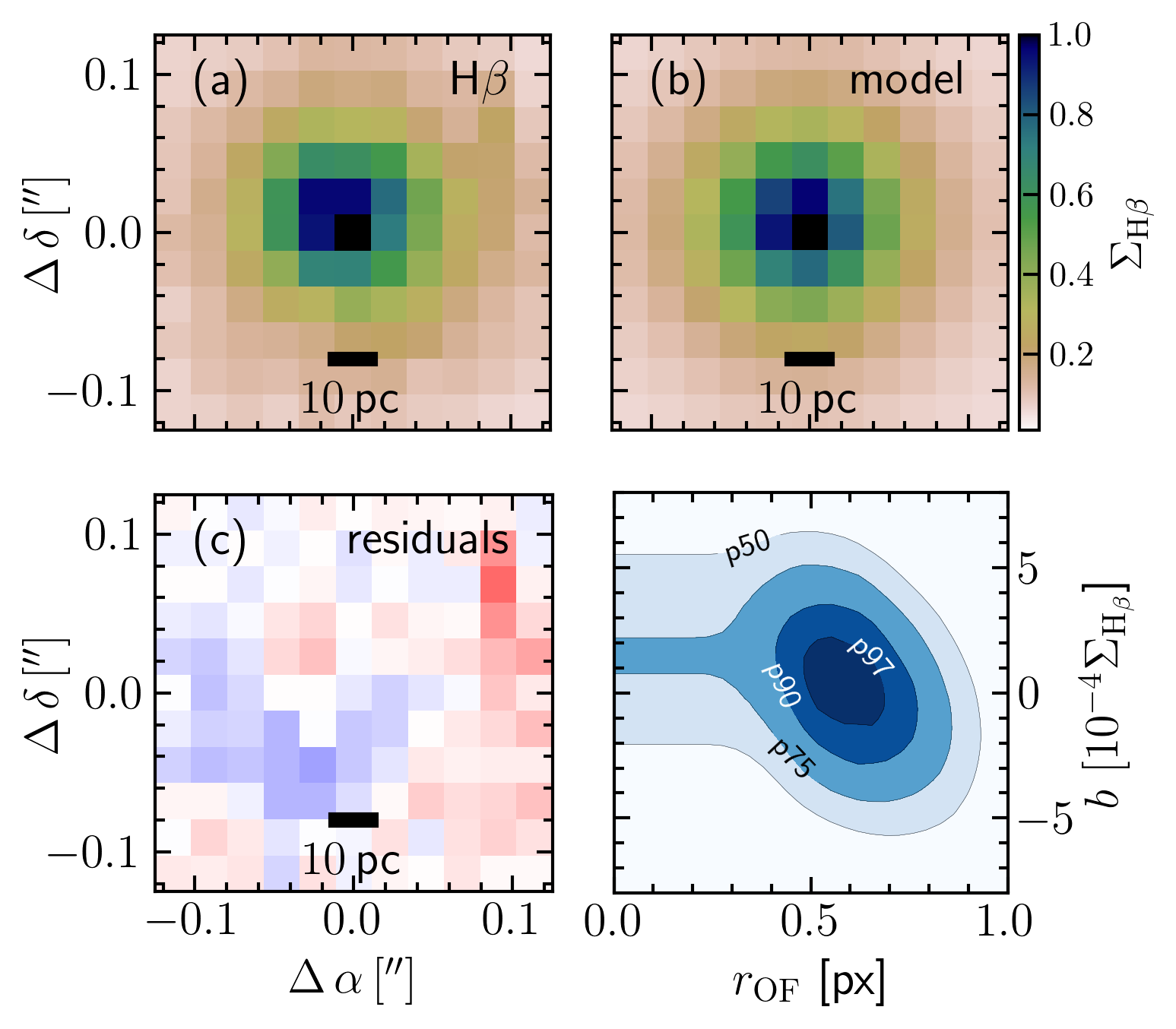}
   \caption{Measuring the size of Mrk~1044's spatially extended H$\beta$ outflow.
    Panel (a) shows the flux map of the H$\beta$ outflowing component which we extracted as described in Paper~I. Panel (b) shows the best-fitting model that consists of the empirical PSF convolved with a Gaussian kernel and a constant background flux. The residuals in panel (c) show that broadened PSF describes the H$\beta$ profile well. Panel (d) shows the $\chi ^2$ distribution in the parameter plane where a global minimum is present around $0.55 \,{\rm px}$ ($4.6\,{\rm pc}$).}
 \label{fig:size}
\end{figure}

H0 is one of the identified ionised gas outflows' components. It is present in the AGN-subtracted host data cube and therefore spatially resolved. In Fig.~\ref{fig:size} we zoom in onto the central $0\farcs1$ of the narrow H$\beta$ surface brightness map where the outflow flux, velocity and velocity dispersion peak. 
To constrain the outflow's location and projected size we fit the surface brightness profile with a model light distribution of an intrinsically extended source. Our model consists of ({\it i}) the empirical PSF extracted at H$\beta$ which we convolved with a Gaussian kernel $r_{\rm OF}$, and ({\it ii}) a constant background flux $b$. To find the best-fitting parameters, we minimise the SNR-weighted $\chi^2$-sum of the H$\beta$ residual flux distribution. 

The best-fit model is shown in the top right panel of Fig.~\ref{fig:size}. It yields an intrinsic size of $r_{\rm OF} = 0.55 \pm 0.10 \,{\rm px}$ ($4.6 \pm 0.6 \,{\rm pc}$) from which we conclude that the source is spatially resolved at a >$3\sigma$ confidence level.
This extended component of the outflow is also present in H$\alpha$ emission. Here, we followed the same method and retrieved an intrinsic projected extent of $0.38 \pm 0.02 \,{\rm px}$ ($3.2 \pm 0.1 \,{\rm pc}$). Both values are consistent within the uncertainties, which suggest that both light profiles stem from the same structure. We measure the centroid of the light distribution which matches the AGN position with an uncertainty of $0.22 \,{\rm px}$ (1.8$\,{\rm pc}$).
Our results are robust against choosing different spectral regions of the broad line windows for the empirical PSF extraction that is described in Paper~I. Furthermore, our results do not change within the uncertainty if we use the AGN-subtracted host emission line maps presented in Paper~I instead of the H$\beta$ and H$\alpha$ surface brightness maps that we retrieve from the spectroastrometric analysis. The former were generated from fitting the AGN-subtracted host emission with a single-Gaussian component, which represents a fundamentally different method from the spectroastrometrically extracted maps.

\section{Geometric expanding shell model}
\label{appendix:shell_model}

\begin{figure}
 \centering
 \resizebox{\hsize}{!}{\includegraphics{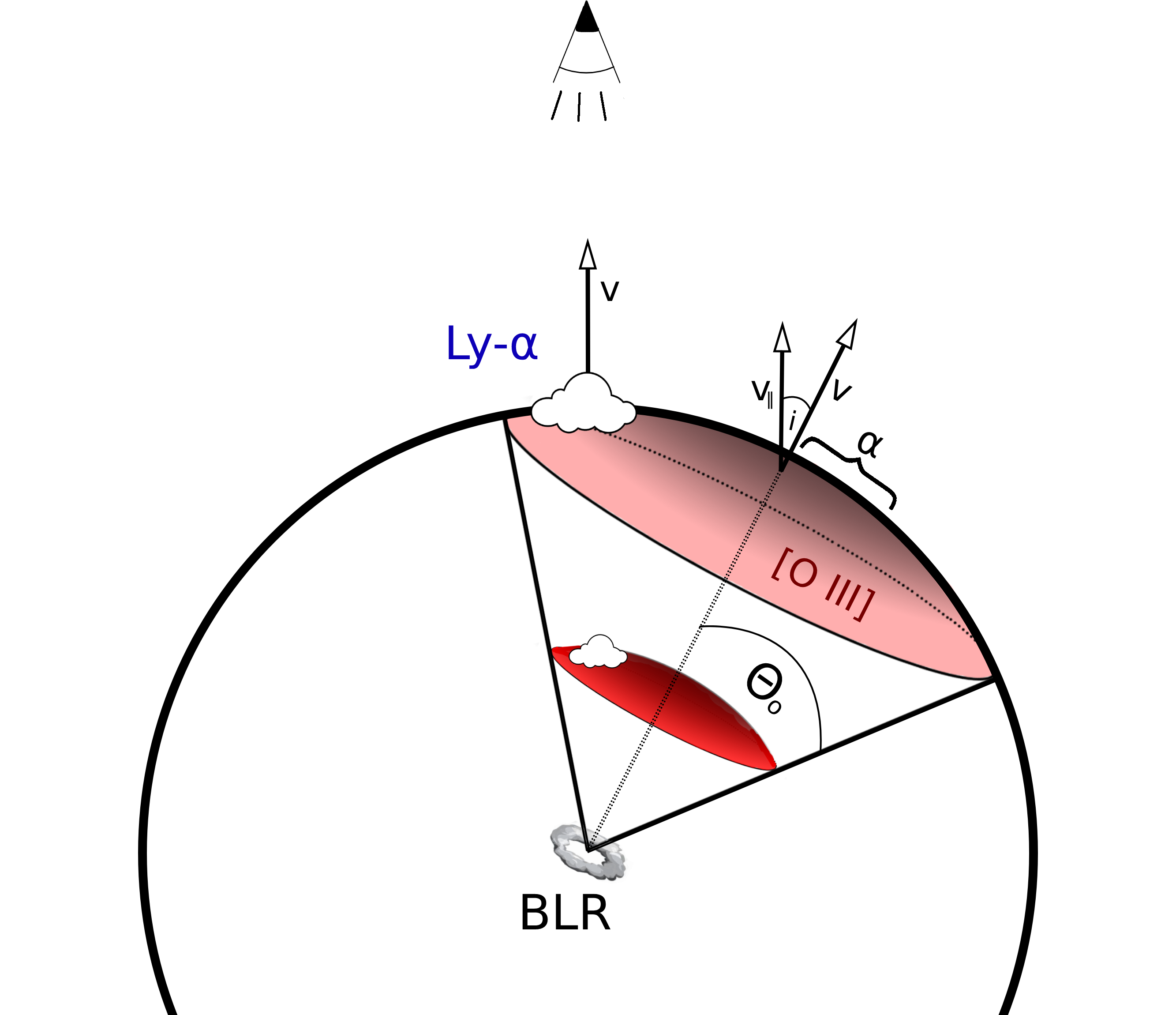}}
 \caption{Illustration of the expanding shell geometry. The Ly-$\alpha$ occurs in the expanding shell, along the LOS towards the BLR with the velocity $v$. The [\ion{O}{iiii}] emission originates from an intrinsically extended but spatially unresolved shell cap with a half-opening angle $\theta_0$, half-light radius $\alpha$ and inclination $i$. The observer measures the integrated flux and thus measures a smaller velocity.}
 \label{fig:shell_model_sketch}
\end{figure}

\begin{figure}
 \centering
 \resizebox{.8\hsize}{!}{\includegraphics{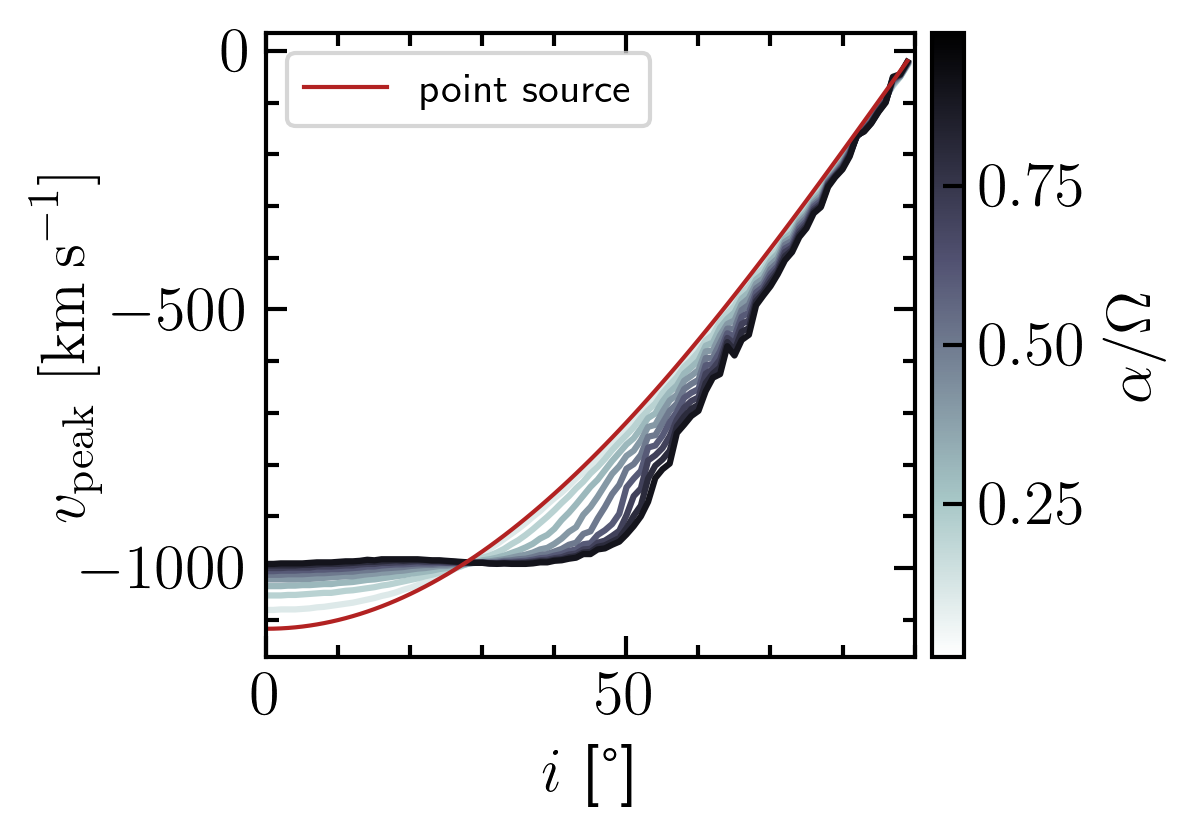}}
 \caption{Dependence of the emission line flux maximum corresponding velocity on inclination. Independent of how concentrated the light profile is ($\alpha/\theta_0$), the geometric model requires high inclinations of the [\ion{O}{iii}]-emitting shell to significantly reduce the peak velocity by a factor of approximately two.
 }
 \label{fig:shell_model_parameters}
\end{figure}

Here, we describe how we use a geometric model of two expanding shells to reproduce the [\ion{O}{iii}] emission line shape. We use this model to constrain the geometric alignment of the unresolved ionised gas outflow relative to the Ly-$\alpha$ absorbers. The implication of the results are discussed in Sect.~\ref{Sect:Geometric_Model}.

\subsection{Constraining the outflow geometry}

\begin{figure*}
 \centering
 \resizebox{\hsize}{!}{\includegraphics{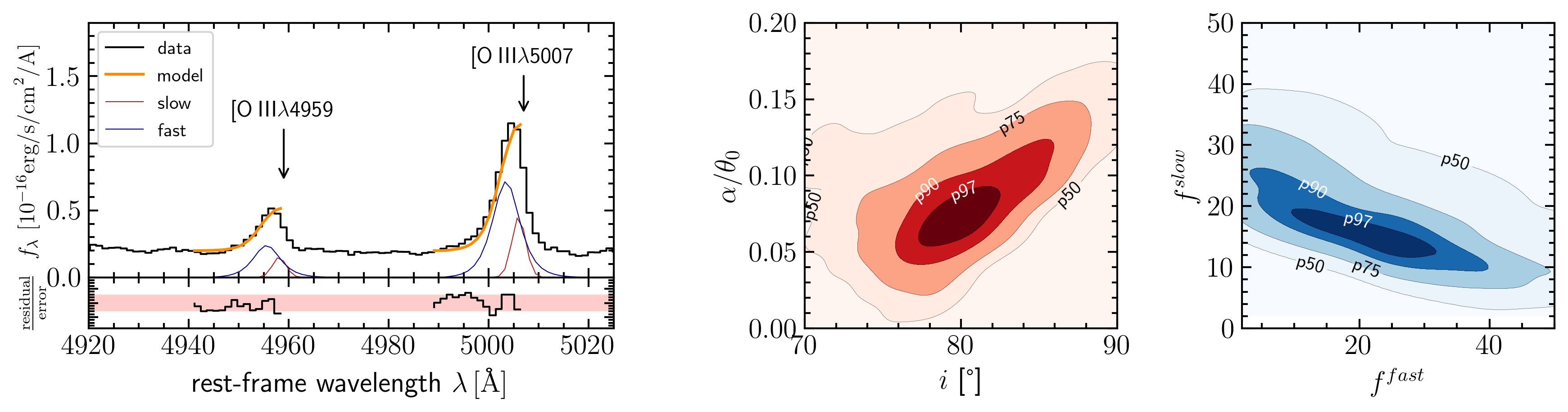}}
 \caption{Best-fitting model for the blue shoulder of Mrk~1044's [\ion{O}{iii}]$\lambda\lambda$4959,5007 emission line.
 The left panel shows the AGN spectrum (black line) around the [\ion{O}{iii}] doublet emission. The best-fit model (orange line) reproduces the line-shape well. Both the {\it slow} (red line) and the {\it fast} shell (blue line) contribute to the asymmetric line shape.
 The panels on the right show percentile regions of the posterior distribution for the geometric parameters (left) and the flux contribution (right) respectively.}
 \label{fig:shell_model_fit}
\end{figure*}

In contrast to the unresolved but intrinsically extended [\ion{O}{iii}] outflow, the UV spectrum only probes the line-of-sight absorption towards the high density BLR clouds. We can link the emitters and absorbers by assuming that a small fraction of the [\ion{O}{iii}]-emitting gas is also responsible for the Ly-$\alpha$ absorption. In this scenario, the systematic velocity factor of $\sim 2$ between [\ion{O}{iii}]-emitters and Ly-$\alpha$-absorbers may be the result of a projection effect.

There exist different descriptions for outflow morphologies including uniformly filled cones \citep[e.g.][]{Cano-Diaz:2012, Cresci:2015, Fiore:2017}, hollow bi-cones \citep[e.g.][]{Muller-Sanchez:2011} or thin shell structure \citep[e.g.][]{Rupke:2005}. In our case, the relatively narrow Ly-$\alpha$ absorption features have low velocity dispersion, which suggests that the absorbing structure is better described by a shell-like shock front scenario.
We therefore assume that the ionised gas is distributed on two concentric shell caps with the AGN located in the origin. The shell caps expand radially with a velocity $v$. We assume that the receding side is obscured by dust.
Both shell caps have a half-opening angle $\theta_0$ that is observationally constrained. \cite{Ricci:2017} applied a torus model to a sample of Swift BAT AGN for which they constrained the median value of the torus half-opening angles to $58 \pm 3 ^{\circ}$. We adopt this value for our model, although varying $\theta_0$ by $\pm 10 ^{\circ}$ does not alter our results qualitatively. 
We assume that luminosity profile on the shell cap is centrally concentrated. We describe it by an exponential luminosity profile that declines radially with a half-light radius $\alpha$ in units of the half-opening angle $\theta_0$. Thus, $\alpha/\theta_0 =1$ means that the half-light radius coincides with the half-opening angle. Furthermore, we include the luminosity ratio between the shells as a free parameter.
An illustration of the geometric model is shown in Fig.~\ref{fig:shell_model_parameters}. The two shells share the geometric parameters $i$ and $\alpha$, but not for the expansion velocity $v$ that is fixed to the measured velocity of the Ly-$\alpha$ absorbers.

\subsection{Predicting the emission line shape}

We use a forward-modelling approach to constrain the parameters of the expanding shell model from the [\ion{O}{iii}]$\lambda\lambda$4959,5007 emission lines. We predict the observed emission line shape with a numerical approach that involves the following steps.

(1) As a first step, we sample the shell cap with 10$^5$ equidistant cells using the Fibonacci sphere algorithm\footnote{using random sampling or the golden spiral method deliver the same results}.

(2) We incline the shell cap using a rotational coordinate transform.

(3) For each of the cells we compute the line-of-sight velocity component and performed a binning in velocity space. This provides us with the predicted emission line shape.

(4) To fit the rest-frame spectrum of Mrk~1044 we transform the distribution into the wavelength space using the expansion velocity of the shell together with the rest-frame wavelengths of the respective ion.
With this method we generate a template of emission lines shapes, computed for a 10$^2  \times 10^2$ parameter grid with $i=[0,90)$ and $\alpha = (0,1]$.

Since we have two emitters and absorbers respectively, our model requires a {\it fast} and a {\it slow} component. 
To illustrate the effect of the geometric parameters in Fig.~\ref{fig:shell_model_parameters}, we show the dependence of the {\it fast} component's peak emission on the inclination of the shell caps. Higher inclinations move the peak of the line closer to the systemic velocity $v_{\rm sys}=0\,{\rm km\:s}^{-1}$, whereas the effect of the central concentration $\alpha/\Omega$ depends on the inclination of the system. The solution for the line shape for the {\it slow} component behaves equivalently.

To find the linear combination of the {\it fast} and {\it slow} component that fits the [\ion{O}{iii}] best, we use a non-negative least square optimisation algorithm. Since we only see the approaching side of the outflow, we only model the blue shoulder of the [\ion{O}{iii}]$\lambda\lambda$4959,5007 in the with corresponding velocity range \mbox{[v$_{\rm L2}$, 0$\,{\rm km\:s}^{-1}$]}.
To get an estimate of the parameters' uncertainties, we use a Monte Carlo approach. From the original AGN spectrum we generate 10$^4$ mock spectra by randomly varying the flux density within the uncertainty. For the resulting parameter distribution, we adopted half of the 16$^{th}$ to 84$^{th}$ percentile range as uncertainty.

\subsection{The best-fitting geometric model}

Using the concentrically expanding shell model, we retrieve an inclination of \mbox{$i = 79.1 \pm 8.4 \,^\circ$}. The best-fit concentration \mbox{$\alpha / \theta_0 = 0.09 \pm 0.04$} indicates a relatively narrow (‘jetted’) luminosity profile.
The results do not change within the uncertainties if the [\ion{O}{iii}] lines are fitted independently. Furthermore, our results do not depend on how we define the ‘blue shoulder’. This means, that slightly adjusting the wavelength-borders around the emission line does not affect the results significantly.

The best-fit geometric parameters have an intuitive explanation. The width of the [\ion{O}{iii}] primarily determines the concentration of the luminosity profile. Since we do not observe a double-peaked emission line, both emission lines have to be close in wavelength. This can only be achieved at high inclinations, as shown in Fig.~\ref{fig:shell_model_parameters}.

Introducing another geometric parameter leads to an over-fitting of the [\ion{O}{iii}] emission lines given their relatively low spectral SNR.
We have tested different parameterisations to describe the luminosity profile.
In particular, using a different parameterisation to describe the geometry of a hollow cone \citep[e.g.][]{Das:2006, Muller-Sanchez:2011,Venturi:2018} requires similar inclinations to explain the [\ion{O}{iii}] line shape.
We conclude that the absence of a single-peaked [\ion{O}{iii}] emission line can only be reproduced by a concentrically expanding shell model if the system is highly inclined with respect to the LOS.

\end{appendix}

\end{document}